\documentclass[10pt,conference]{IEEEtran}
\usepackage{amsmath,amsfonts}
\usepackage{algorithmic}\usepackage[export]{adjustbox}
\usepackage{graphicx}
\usepackage{textcomp}
\usepackage{rotating}
\usepackage{array}
\usepackage{xcolor}
\usepackage{tikz}
\usepackage{makecell}
\usepackage{multirow}
\usepackage{dblfloatfix}
\usepackage[export]{adjustbox}
\usepackage{etoolbox}
\usepackage{url}
\usepackage{caption,setspace}
\def\BibTeX{{\rm B\kern-.05em{\sc i\kern-.025em b}\kern-.08em
    T\kern-.1667em\lower.7ex\hbox{E}\kern-.125emX}}
\begin{document}

\newcommand*\circled[1]{\tikz[baseline=(char.base)]{
		\node[shape=circle,draw,inner sep=0.8pt] (char) {#1};}}
	
%\setcopyright{acmcopyright}

\title{Label Smoothing Improves Neural\\Source Code Summarization
}

\author{\IEEEauthorblockN{Sakib Haque,
		Aakash Bansal, and
		Collin McMillan\\
		\IEEEauthorblockA{Department of Computer Science\\
			University of Notre Dame, Notre Dame, IN, USA\\
			Email: \{shaque, abansal1, cmc\}@nd.edu}}}

\maketitle

%\IEEEtitleabstractindextext{
\begin{abstract}
Label smoothing is a regularization technique for neural networks.  Normally neural models are trained to an output distribution that is a vector with a single 1 for the correct prediction, and 0 for all other elements.  Label smoothing converts the correct prediction location to something slightly less than 1, then distributes the remainder to the other elements such that they are slightly greater than 0.  A conceptual explanation behind label smoothing is that it helps prevent a neural model from becoming ``overconfident'' by forcing it to consider alternatives, even if only slightly.  Label smoothing has been shown to help several areas of language generation, yet typically requires considerable tuning and testing to achieve the optimal results.  This tuning and testing has not been reported for neural source code summarization -- a growing research area in software engineering that seeks to generate natural language descriptions of source code behavior.  In this paper, we demonstrate the effect of label smoothing on several baselines in neural code summarization, and conduct an experiment to find good parameters for label smoothing and make recommendations for its use.
\end{abstract}

\begin{IEEEkeywords}
	Source code summarization, automatic documentation generation, label smoothing, optimization
\end{IEEEkeywords}
%}

%\keywords{source code summarization, automatic documentation generation, label smoothing, optimization}

%\maketitle

\section{Introduction}

The backbone of much software documentation is the ``source code summary''~\cite{haiduc2010use, sridhara2010towards, kramer1999api}.  
A summary is a short description in natural language that provides high level information about the purpose and behavior of the low level details implemented in source code.  
The idea is that a programmer can understand the functionality of a section of code by reading the corresponding summary without reading the code itself~\cite{forward2002relevance, xia2017measuring}.  
The expense of writing these summaries by hand has long made automated code summarization a ``holy grail''~\cite{leclair2019neural} of software engineering research~\cite{zhao2020survey}.  
The dream is that programmers could read code documentation even if other programmers did not write any.

The state-of-the-art in code summarization research depends on the neural encoder-decoder architecture~\cite{leclair2019neural, zugner2021languageagnostic, hu2018deep}.  
Basically the idea is that an encoder forms a representation of source code in a vector space, while the decoder forms a representation of the summary in a different vector space. 
With sufficient training data (usually millions of samples~\cite{allamanis2018survey, leclair2019recommendations}), another part of the model learns to connect features in one space to the other and can be used to predict output summaries for arbitrary input source code.  
Neural designs based on the encoder-decoder model have almost completely supplanted earlier template- and heuristic-based approaches~\cite{zhao2020survey}.

These neural models generate summaries one word at a time (usually using the teacher forcing procedure~\cite{toomarian1992learning} in a seq2seq-like design~\cite{sutskever2014sequence, zhao2020survey}).  In a nutshell, the model is shown the source code and a start of sequence token for the desired output summary.  The model then predicts the first word of the output summary.  Then the model is shown the source code and the start of sequence token plus the first predicted word, then predicts the second word.  This process continues until the model predicts an end of sequence token.  The point is that the model is tasked with predicting a single word several times -- not the whole output summary at once.  Each individual word prediction is a weakpoint: if the model makes an error, the subsequent predictions are also likely to be wrong because they depend on the previous one~\cite{haque2021action}.  If the model is ``overconfident'' in predicting some words, it may develop a tendency to miss rarer words and produce repetitive, dull outputs~\cite{jiang2018sequence}.

A solution proffered in several areas of natural language generation is \emph{label smoothing}~\cite{muller2019does, lukasik2020does, yuan2020revisiting}.  Label smoothing seeks to make the model ``less confident'' in each prediction by altering the target output.  The neural model's output is a predicted probability distribution over the entire vocabulary of words.  If the vocabulary has 10,000 words, then the output is technically a 10,000-length vector, in which the position of the correct word will hopefully have the highest value.  During training, the target vector would have a 1 in the position of the correct word, and a 0 in the other 9,999 positions.  What label smoothing does is reduce the value in the correct position slightly, say to 0.95, then spread the remainder over the rest of the distribution, so all other positions would be e.g., (0.05/9999).  While the mechanism by which label smoothing works is not fully understood~\cite{lukasik2020does}, different empirical studies have repeatedly shown it to be effective~\cite{muller2019does}.

In this paper, we present an empirical study on label smoothing for neural source code summarization.  Label smoothing requires considerable tuning to achieve optimal results in different domains.  While it is reasonable to hypothesize that label smoothing would improve neural code summarization given the similarities between neural code summarization and other natural language generation technologies, the hypothesis has not been tested, and effective parameters and procedures have not been established.  We conduct an experiment of label smoothing for several baselines from the source code summarization literature, to serve guide for future researchers.

\section{Background \& Related Work}

In this section, we discuss the key background ideas and supporting work related to this paper, namely source code summarization, neural encoder-decoder architecture, and label smoothing.

\vspace{-0.1cm}
\subsection{Source Code Summarization}
\label{sec:backgroundscs}
Early work in source code summarization included heuristic-based approaches~\cite{sridhara2011automatically, mcburney2014automatic}.
These models relied on techniques of Information Retrieval (IR) to extract salient words from source code~\cite{de2012using, rodeghero2014improving}.
These words were then put into manually-defined templates to produce meaningful sentences~\cite{moreno2014automatic, zhang2016towards}.

\begin{table}[!b]
	%\begin{table}[t!]
	\vspace{-.5cm}
	{\small
		\begin{tabular}{p{3.9cm}p{0.4cm}p{0.4cm}p{0.4cm}}
			& N         & S          & C            \\
			%\textcolor{white}{*}Oda~\emph{et al.}~(2015)~\cite{oda2015learning}						& x &   & \\
			\textcolor{white}{*}Iyer~\emph{et al.}~(2016)~\cite{iyer2016summarizing}				& x &   & \\
			\textcolor{white}{*}Loyola~\emph{et al.}~(2017)~\cite{loyola2017neural}					& x &   & \\
			%\textcolor{white}{*}Lu~\emph{et al.}~(2017)~\cite{lu2017learning}						& x &   & \\
			\textcolor{white}{*}Jiang~\emph{et al.}~(2017)~\cite{jiang2017automatically}			& x &   & \\
			\textcolor{white}{*}Hu~\emph{et al.}~(2018)~\cite{hu2018summarizing}					& x &   & \\
			\textcolor{white}{*}Hu~\emph{et al.}~(2018)~\cite{hu2018deep}							& x & x & \\
			\textcolor{white}{*}Allamanis~\emph{et al.}~(2018)~\cite{allamanis2018learning}			& x & x & \\
			\textcolor{white}{*}Alon~\emph{et al.}~(2019)~\cite{alon2019code2seq}					& x & x & \\
			\textcolor{white}{*}Gao~\emph{et al.}~(2019)~\cite{gao2019neural}						& x &   & \\
			\textcolor{white}{*}LeClair~\emph{et al.}~(2019)~\cite{leclair2019neural}				& x & x & \\
			\textcolor{white}{*}Fernandes~\emph{et al.}~(2019)~\cite{fernandes2018structured}		& x & x & \\
			\textcolor{white}{*}Haque~\emph{et al.}~(2020)~\cite{haque2020improved}					& x & x & x \\
			\textcolor{white}{*}Haldar~\emph{et al.}~(2020)~\cite{haldar2020multi}					& x & x & \\
			\textcolor{white}{*}LeClair~\emph{et al.}~(2020)~\cite{leclair2020improved}				& x & x & \\
			\textcolor{white}{*}Ahmad~\emph{et al.}~(2020)~\cite{ahmad2020transformer}				& x & x & \\
			\textcolor{white}{*}Z\"{u}gner~\emph{et al.}~(2021)~\cite{zugner2021languageagnostic}	& x & x & \\
			\textcolor{white}{*}Liu~\emph{et al.}~(2021)~\cite{liu2021retrievalaugmented}			& x & x & \\
			\textcolor{white}{*}LeClair~\emph{et al.}~(2021)~\cite{leclair2021ensemble}				& x & x & \\
			\textcolor{white}{*}Gao~\emph{et al.}~(2021)~\cite{gao2021code}							& x & x & \\
			\textcolor{white}{*}Wang~\emph{et al.}~(2021)~\cite{wang2021cocosum}					& x & x & \\
			\textcolor{white}{*}Bansal~\emph{et al.}~(2021)~\cite{bansal2021project}				& x & x & x \\
			\textcolor{white}{*}Gong~\emph{et al.}~(2022)~\cite{gong2022source}						& x & x & 
		\end{tabular}
	}
	\vspace{0.2cm}
	\caption{\small{Selection of closely-related, peer-reviewed works that use neural network based architecture.  Column $N$ indicates Neural Network inspired solutions. $S$ indicates that structural data (various representations of AST) is used. $C$ indicated publications that incorporate contextual information.}}
	\label{tab:screlated}
	%\end{table}
\end{table}

The landscape changed with the advent of deep learning.
The explosion of data-driven models in the mid-2010's and their state-of-the-art performance in various natural language processing (NLP) tasks inspired researchers to use them to automatically generate source code summaries.
Iyer~\emph{et al.}~\cite{iyer2016summarizing} was one of the earliest to use such models for code summarization.
Since then, the field has embraced this sequence-to-sequence (seq2seq) architecture as the standard architecture to generate comments.
Figure~\ref{tab:screlated} lists some of the most prominent papers that use some modified version of an attentional neural encoder-decoder model to generate code summaries.
This list is not exhaustive, but only includes peer-reviewed papers where a new idea was first introduced.

The table shows that the first generation of these data-driven approaches (marked only in column $N$) only looked at the source code tokens to generate comments.
Later, Hu~\emph{et al.}~\cite{hu2018deep} and LeClair~\emph{et al.}~\cite{leclair2019neural} noted the importance of including structural information about the source code by using the Abstract Syntax Tree (AST).
They both flattened the AST into sequential tokens, but incorporated these tokens in different ways in their model.
Other research papers soon explored various ways of capturing these structural information from the AST~\cite{allamanis2018learning, alon2019code2seq, leclair2020improved, fernandes2018structured}.
At the same time, a parallel research track delved into incorporating contextual information.
This context encompasses API calls to learn the mapping between API sequences and natural language description~\cite{hu2018summarizing} as well as other functions in the file to provide supporting information for the code~\cite{haque2020improved}.
Bansal~\emph{et al.} further expanded the latter idea by including project context information~\cite{bansal2021project}.
Recently, some research has been dedicated into using transformer models~\cite{vaswani2017attention} for this task.
Ahmad~\emph{et al.}~\cite{ahmad2020transformer} used pairwise relationship between tokens to capture their mutual interaction beyond positional encoding while Gong~\emph{et al.}~\cite{gong2022source} introduces another layer in the encoder side of the transformer for the AST.

This paper aims to improve the performance of these data-driven models by exploring how label smoothing can enhance their performance.
We show how the performance of different established baselines improves using label smoothing.
Note that we do not seek to compete with any one code summarization approach -- our aim is to benefit all approaches.

\vspace{-0.1cm}
\subsection{Neural Encoder-Decoder Architecture}
\label{sec:backgroundnmt}
The neural encoder-decoder architecture is the backbone of almost all current code summarization approaches~\cite{haque2021action}.
These models are borrowed from the field of Neural Machine Translation (NMT).
Essentially, these models have 2 parallel components: the encoder and the decoder.
The encoder takes words (tokens) from the input language (e.g. English for NMT/Java for source code summarization) and represents them into fixed length vectors.
The decoder takes this vector representation and translates them to the target language (e.g. Spanish for NMT/English for source code summarization).

The standard setup for these encoder-decoder models use recurrent layers on both sides~\cite{sutskever2011generating}.
% Cite original LSTM paper here
These recurrent layers are typically a sequence of LSTM~\cite{hochreiter1997lstm} or GRU~\cite{cho2014learning} cells.
Information propagates through these layers, one token per cell.
Each cell in the encoder layer not only receives the corresponding token, but also the vector representation of all the tokens that came before.
The first cell in decoder layer receives the final vector representation of the entire input sequence~\cite{sutskever2014sequence}.
Consecutive cells in the decoder layer receive a vector that encapsulates not only the entire input sequence but also the output so far.

While this architecture has existed for some time, Bahdanau~\emph{et al.}~\cite{bahdanau2014neural} enhanced its performance by introducing attention in 2014. 
The intuition behind attention is that some tokens in the input sequence are more important than others when trying to generate specific predictions.
The goal of attention is to map the relative importance of each input token to every output prediction.
It does this by computing the similarity between the input sequence and the output sequence so far to identify the important features to predict the next word.
The two most common attention functions are additive and multiplicative (dot-product based).
In all the models evaluated in this paper, we use multiplicative attention.

The dot-product based attention inspired a new generation of neural encoder-decoder architecture called Transformers, that eschews the sequential nature of these models.
First introduced by Vaswani~\emph{et al.}~\cite{vaswani2017attention}, these models replace the RNN cells with a self-attention layer (discussed in greater detail in section~\ref{sec:baselines}) followed by a fully connected feed-forward layer.
Without the recurrent layer to propagate sequential information, transformers introduce a positional encoding before the encoder and decoder respectively to capture the order of the sequence.
This allows for parallel processing of tokens, making them faster.
These models have also been shown to better capture long-range relationships~\cite{vaswani2017attention}.

%\vspace{-0.1cm}
\subsection{Label Smoothing}
\label{sec:backgroundls}
Label smoothing is a regularization technique used in neural networks to improve generalization.
It was first introduced by Szegedy~\emph{et al.}~\cite{szegedy2015rethinking} for image classification, although it has since been shown to improve performance for many other deep learning tasks, including NMT~\cite{muller2019does}.
Essentially, label smoothing involves introducing some uncertainty in the training data to prevent over-fitting~\cite{lienen2021label, lukasik2020does, zhang2021delving}.

Most Natural Language Generation (NLG) models use categorical cross-entropy loss to calculate the error between target tokens and predicted tokens.
Minimizing this loss function means reducing the error between the target and predicted tokens, thus improving model performance.
The target distribution ($t(k)$) for a neural network is a Dirac delta function: 
\[t(k) = \delta_{k, y}\] 
where $k$ is the predicted output and $y$ is the target output and $t(k)$ is 1 when $k=y$ and 0 for all other $k$.
In practice, we use a one-hot vector to represent the Dirac delta function.
The size of the vector is the number of tokens in the output vocabulary ($N_t$), with all elements set to 0 except for the index of the target token, which is set to 1.
We can therefore read it as a probability distribution that is 1 for the correct target word and 0 for all other words in the vocabulary.
This is undesirable because it makes the model overconfident about certain predictions.

With label smoothing, we encourage the model to be ``less confident.''
We achieve this, in essence, by taking some probability, $\epsilon$, off the top of the target token, $y$ and uniformly distributing over the rest of the vocabulary.
This probability, $\epsilon$ where $\epsilon\in(0,1]$, is the key parameter that determines the extent to which label smoothing will affect model performance.
We represent this new distribution as:
\[t(k) = (1-\epsilon)\delta_{k, y}+(1-\delta_{k, y})\frac{\epsilon}{N_t-1}\]
where $t(k)$ is now $(1-\epsilon)$ for $k=y$ and $\frac{\epsilon}{N_t-1}$ for $k\neq y$.

The current research frontier in source code summarization lies in enhancing the neural networks to generate better comments.
This paper lies at the heart of this frontier by not only quantifying the regularization effect of these models with label smoothing but also identifying the best configuration to maximize model performance.
\section{Experimental Design}
In this section, we discuss the design of our experiment.
This includes the research objectives of this paper, the datasets we use to perform the experiments, the models we use to train the data, the metrics we use to evaluate model performance, and any threats to validity we might have.

\vspace{-0.2cm}
\subsection{Research Questions}
\label{sec:rqs}
Our research objective is to evaluate the extent to which label smoothing improves source code summarization models, and the configurations that maximize this performance improvement. 
To this end, we ask the following research questions:
\begin{itemize}
	\item[\textbf{RQ1}] What is the effect of label smoothing on the overall performance of recently-published source code summarization baselines in terms of BLEU, METEOR and USE+c scores?
	\item[\textbf{RQ2}] What is the best label smoothing configuration that maximizes model performance and how is this configuration affected by the output vocabulary size?
	\item[\textbf{RQ3}] What is the effect of label smoothing on the diversity of target vocabulary and how does this affect  performance?
\end{itemize}

The rationale behind RQ1 is that label smoothing is not a commonly used regularization technique in automatic source code summarization, despite evidence of success in other NLG applications.
Many models have been proposed for source code summarization, but to the best of our knowledge, none use label smoothing as a regularizer.
While there is evidence that label smoothing will improve the performance of these models, there has been no extensive research to demonstrate this.
Our goal with this RQ is to quantify how including label smoothing to train these models can change their performance.
We use three different metrics to verify our findings: BLEU, METEOR and USE+c.
These metrics are further discussed in Section~\ref{sec:metrics}

The rationale behind RQ2 is that our application of label smoothing has underlying hyperparameters that require tuning.
As noted earlier, $\epsilon\in (0,1]$.
But the research literature is unclear about what the optimal values are for these hyperparameters, especially in the problem of code summarization.
Additionally, we also do not know how the size of the output vocabulary ($N_t$) affects model performance with this configuration.
The larger the vocabulary size, the smaller the smoothed probability per output token ($\frac{\epsilon}{N_t}$).
Our goal with this RQ is to identify a suitable value of $\epsilon$. % that increases BLEU, METEOR and USE+c scores the most.
Furthermore, we aim to understand the relationship of this probability $\epsilon$ with the size of output vocabulary, $N_t$.

The rationale behind RQ3 is that there are competing intuitions as to how label smoothing affects model performance.
As noted in Section~\ref{sec:backgroundls}, label smoothing artificially introduces uncertainty in the one-hot output vector.
On one hand, the introduction of uniform uncertainty across the output vector could exacerbate the problem of label noise~\cite{xie2016disturblabel}.
However, this uncertainty makes the model less confident about its predictions and allows it to consider rare words, thus reducing the problem of class imbalance in source code summarization~\cite{haque2021action}.
On the other hand, preventing overconfidence could mitigate label noise by improving model generalization, thus focusing on more commonly occurring words~\cite{lukasik2020does}.
Our goal with this RQ is to study which of these competing intuitions is borne out in practice for the problem of source code summarization.

\subsection{Datasets}
We use two datasets in this paper: one is Java and the other is C/C++.
The Java dataset, first published by LeClair~\emph{et al.}, introduce some of the best practices for developing a dataset for source code summarization that are now a standard among the research community~\cite{leclair2019recommendations}.
It consists of 2.1m methods from more than 28k projects.
The training, validation and test set are split by projects to prevent data from train set to leak into the test set by virtue of being in the same project.
This dataset has since been used in many peer-reviewed publications~\cite{haque2020improved, leclair2020improved, stapleton2020human, bansal2021project, xie2021exploiting} and new additions has since been made to it, including context tokenization.
We use a filtered version of this dataset, with 1.9m functions, published by Bansal~\emph{et al.} that remove code clones in accordance with recommendations by Allamanis~\emph{et al.}~\cite{allamanis2019adverse}.

The C/C++ dataset was first published by Haque~\emph{et al.}~\cite{haque2021action} following an extraction model proposed by Eberhart~\emph{et al.}~\cite{eberhart2019automatically} to adhere to the idiosyncrasies of C/C++, while maintaining the same strict standards proposed by LeClair~\emph{et al.}~\cite{leclair2019recommendations}.
It consists of 1.1m methods from more than 33k projects.
% We perform similar filtration on this dataset to remove code clones~\cite{allamanis2019adverse, bansal2021project}.

Additionally, we extract individual statements on the top 10\% largest methods from the Java dataset (Java-q90) and the top-25\% largest methods from the C/C++ dataset (C/C++-q75).
While this filtration reduces the size of the dataset, it eliminates simple getters/setters and other small functions whose comments are easy to predict.
The remaining functions have more statements, which is more challenging and representative of real world use-case scenario.
We chose top 10\% for Java and 25\% for C/C++ to keep the average size and number of subroutines similar for both datasets.
%We call these the Java-q90 dataset and C/C++-q75.

For RQ3, we use the method outlined by Haque~\emph{et al.} to convert both the Java-q90 and C/C++-q75 dataset to action word prediction dataset~\cite{haque2021action}.
We extract action words from comments and stem them to reduce vocabulary size (e.g. to ensure that `delete', `deletes', `deleted', `deleting' are classified as the same action word).

\vspace{-0.2cm}
\subsection{Baselines}
\label{sec:baselines}
%To demonstrate the effect of label smoothing, we explore a wide range of architectures in this paper.
We use six baselines in this paper.
We chose each baseline because it represents a family of similar approaches or is a well-cited approach used as a baseline in many papers.
%The baselines use a range of inputs.
%Some use source code only, other use the AST and others use a combination.
%They also vary in inputs, spanning from code-only to code+AST to code+AST+context.
%Due to space limitations, we only discuss them briefly here.
%Detailed implementation of each model can be found in our online appendix (Section~\ref{sec:appendix}).

\textbf{attendgru} This baseline is a simple unidirectional RNN-based attentional neural encoder-decoder architecture.
It takes only source code tokens as encoder input and English comment as decoder input.
It was first introduced by Iyer~\emph{et al.}~\cite{iyer2016summarizing} as an off-the-shelf NMT/NLG approach to generate source code summaries.
For our implementation, we use GRUs instead of LSTMs because they are much faster while providing comparable performance~\cite{leclair2019neural}.

\textbf{transformer} This baseline is another simple encoder-decoder architecture, but it replaces the recurrent layers with stacked muti-head attention layers~\cite{vaswani2017attention}.
As mentioned in Section~\ref{sec:backgroundnmt}, transformers introduce a position embedding layer that captures the sequential order of tokens which allows the multi-head attention layer to process the entire sequence at the same time.
On the encoder side, the multi-head attention layer computes dot-product based self-attention on the source code tokens. %before passing this to the feed-forward layer.
On the decoder side, there are two multi-head attention layers: a masked multi-head attention layer that computes self-attention on the comment tokens followed by a regular multi-head attention layer that computes attention between the encoder and the masked attention layer.
%The output of this layer is then passed through a feed forward layer to generate output.
%Ahmad~\emph{et al.} shows that transformer based models perform well on source code summarization task~\cite{ahmad2020transformer}.

\textbf{ast-attendgru} This baseline is an enhancement over the attendgru model by including AST information on the encoder side along with source code tokens.
This idea was first proposed by Hu~\emph{et al}~\cite{hu2018deep} who designed a Structure-Based Traversal (SBT) algorithm to flatten the AST and include the source code and AST input together in the encoder.
LeClair~\emph{et al.} improved this model by incorporating this flat AST on a separate recurrent layer and concatenating this encoder output with the original encoder output with just the code tokens~\cite{leclair2019neural}.
For our evaluation, we use the implementation by LeClair~\emph{et al.} because it's more recent and performs better.

\textbf{code2seq} This baseline, proposed by Alon~\emph{et al.}~\cite{alon2019code2seq} is similar to ast-attendgru as it also takes both source code tokens and AST as input.
However, instead of flattening the AST, they encode pairwise paths between nodes in the AST.
Then, they randomly select a subset of these paths as training input.
Randomly selecting paths prevents over-fitting while keeping the model size reasonable.
To reduce architectural variations and manage resource constraints, we set the number of paths explored to 100.

\textbf{codegnngru} This baseline provides another different technique for representing AST along the encoder.
Proposed by LeClair~\emph{et al.}, it uses Convolutional Graph Neural Networks (ConvGNN) to process the AST input~\cite{leclair2020improved}.
They pass the AST nodes through an embedding layer.
The output of this layer along with the AST edge data is then input to the ConvGNN.
%They pass the AST nodes and AST edge data to the ConvGNN.
For each node, ConvGNN adds the neighboring node to it's current node during a hop.
The model shows best performance for 2 hops; each AST node adds the neighboring node twice, thus propagating information between nodes that are separated by 2 hops.
The output of the ConvGNN layer is then passed to a recurrent layer before the result is concatenated with the output of the encoder that processes the source code tokens. 

\textbf{ast-attendgru-fc} This baseline improves upon the ast-attendgru model by including file context information on the encoder side along with source code tokens and AST.
Haque~\emph{et al.} applied the concept of using contextual information from other functions in the same file to a few different baselines~\cite{haque2020improved}.
They introduced a new encoder to process other functions in the file.
All encoder outputs are combined before passing it to the decoder for prediction.
While their results showed improvements in all baselines using file context, ast-attendgru-fc was the highest performing model.

\begin{figure*}[!b]
	%\vspace{0.5cm}
	\minipage{0.32\textwidth}
	\vspace{-0.6cm}
	\includegraphics[width=\linewidth, height=6.5cm]{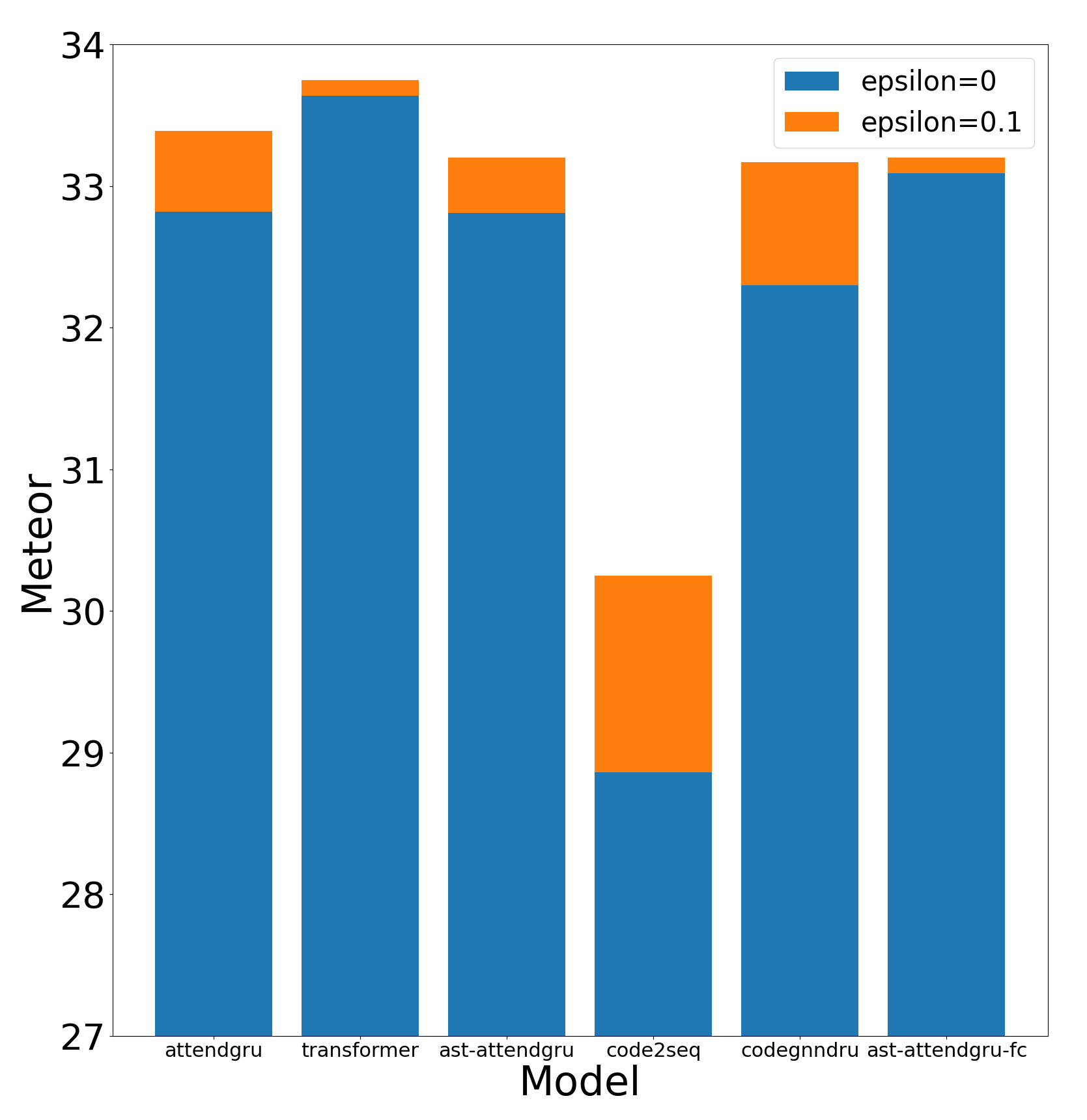}
	%\caption{A really Awesome Image}\label{fig:rq1q90}
	\endminipage\hfill
	%\vspace{0.5cm}
	\minipage{0.32\textwidth}
	\vspace{-0.4cm}
	\includegraphics[width=\linewidth, height=6.5cm]{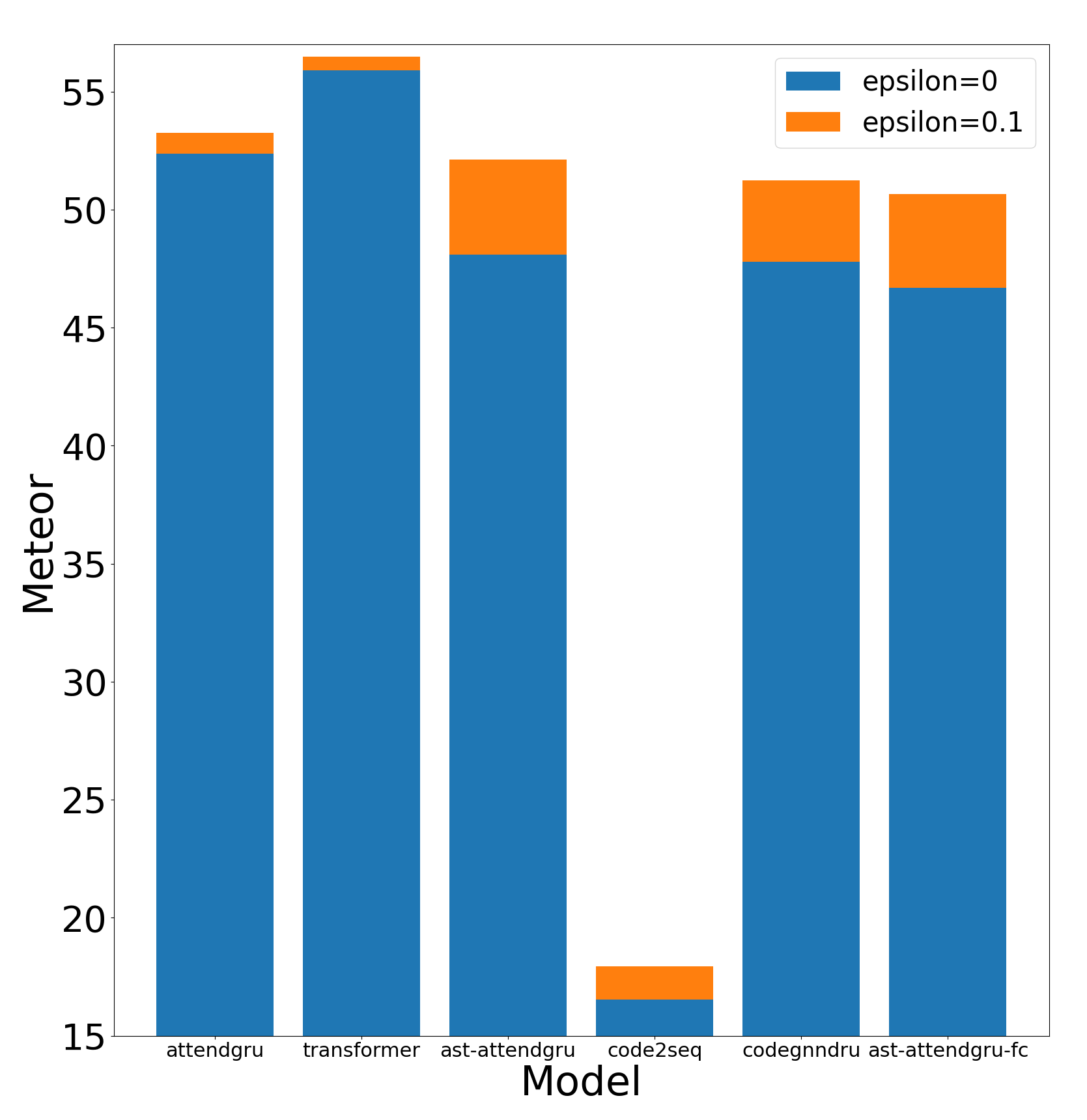}
	%\caption{A really Awesome Image}\label{fig:rq1q75}
	\endminipage\hfill
	%\vspace{0.5cm}
	\minipage{0.32\textwidth}
	\vspace{-0.4cm}
	\includegraphics[width=\linewidth, height=6.5cm]{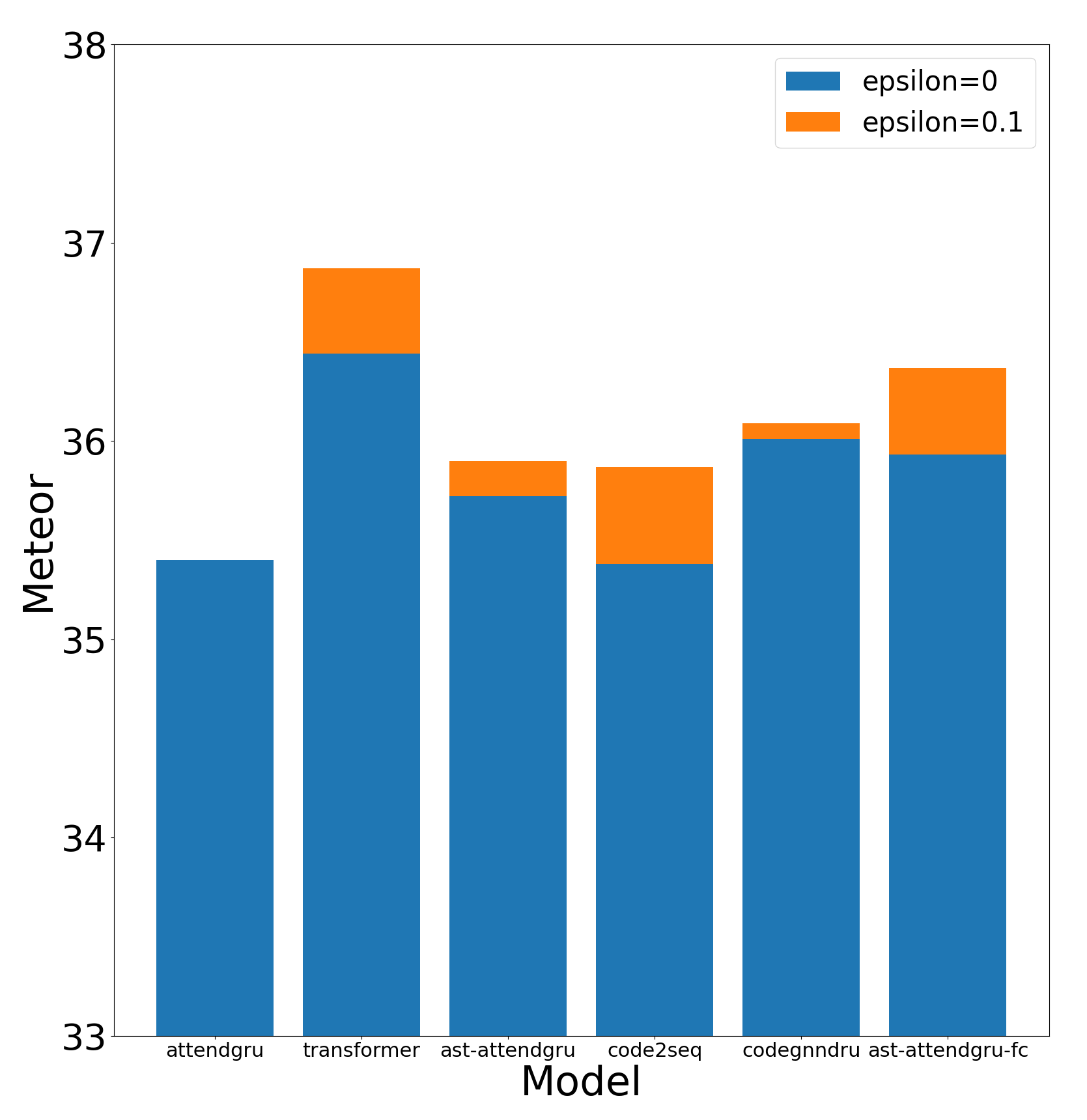}
	\endminipage
	\captionsetup{justification=centering}
	\vspace{-0.2cm}
	\caption{Comparison of baselines with and without label smoothing ($\epsilon$=0.10). Blue indicates baseline aggregate METEOR score. Orange indicates increase in METEOR score for identical model with label smoothing added. Note y-axis starts at 27 METEOR for the Java-q90 dataset (left), 15 for the C/C++-q75 dataset (center) and 33 for the full Java dataset figure (right).}\label{fig:rq1}
	%\vspace{-0.1cm}
\end{figure*}

\vspace{-0.2cm}
\subsection{Metrics}
\label{sec:metrics}
We evaluate these baselines using three different evaluation metrics: BLEU, METEOR and USE+c.
Additionally, for RQ3, we use precision, recall and F1-score to evaluate action word prediction~\cite{powers2011evaluation}.
BLEU and METEOR are n-gram matching metrics while USE+c is a semantic similarity metric.
We use BLEU because it is the most popular evaluation metric for source code summarization.
BLEU is a precision based measure of N-gram overlap~\cite{papineni2002bleu}.
It was first introduced by Papineni~\emph{et al.} in 2002 for automatic evaluation of machine translation tasks.
It compares n-grams in the predicted and target summaries.
Typical implementations of BLEU scores set the range of n from 1 to 4.
An average BLEU score is then computed by combining these individual n-gram scores using predetermined weights.
Most code summarization models use a uniform weight distribution of 0.25 for each n, so we adhere to the same weights.

%We include BLEU score to allow a commensurable comparison between our work and related work.
Recent studies~\cite{stapleton2020human} have shown that BLEU does not correlate well with human judgement of source code comments. 
Roy~\emph{et al.}~\cite{roy2021reassessing} and Haque~\emph{et al.}~\cite{haque2022semantic} have proposed METEOR and USE+c as alternatives that better correlate with human evaluation.
METEOR~\cite{banerjee2005meteor} was introduced in 2005 to address the concerns of using BLEU~\cite{papineni2002bleu} or ROUGE~\cite{lin2004rouge}.
It combines n-gram precision and n-gram recall by taking their harmonic mean to compute a measure of similarity.
Like BLEU, METEOR tries to find exact n-gram matches.
If however, an exact match is not to be found, it performs a second pass to match word stems and finally a third pass, to match word synonyms.% using the wordnet corpora.
METEOR does not have a long history of being used in source code summarization.
However, it was recently shown to better reflect human quality assessment of generated summaries than BLEU or ROUGE, so we report this score.

USE+c~\cite{haque2022semantic} is a new evaluation metric proposed for source code summarization.
It differs from BLEU and METEOR because it does not focus on n-gram matching.
Instead, it uses the pre-trained Universal Sentence Encoder (USE)~\cite{cer2018universal} to compute a vector representation of both target and predicted sentence.
Note that USE trains stacked transformer encoders to compute context aware representation of words in a sentence before turning them into fixed length embeddings.
USE+c then computes the cosine similarity between these fixed length vectors.
A shortcoming of USE+c is that it does not include a brevity penalty.
Unlike BLEU or METEOR, it is also difficult to explain.
However, a recent survey by Haque~\emph{et al.}~\cite{haque2022semantic} showed that USE+c is better correlated with human ratings in terms of similarity, accuracy and completeness than most n-gram based metrics (including BLEU and METEOR).
Therefore, we report USE+c as an additional metric.

%Additionally, for RQ3, we use precision, recall and F1-score to evaluate action word prediction~\cite{powers2011evaluation}.
%This is because action word prediction is a multi-class classification problem, while BLEU, METEOR and USE+c are more suited for full sentence prediction.
%We also use confusion matrices to demonstrate model performance across the most commonly occurring words.

\vspace{-0.1cm}
\subsection{Threats to Validity}

Like any study, our experiment carries threats to validity.
One key threat affecting this paper is the datasets we use.
We aim to mitigate the first threat by using two peer-reviewed datasets with millions of examples in two different and widely-used programming languages.
While we perform some additional filtration of the datasets, we still train our models on a couple hundred thousand functions.
%These datasets have been vetted by the research community.
However, it is still possible the results may not be representative to all datasets.
In this regard, we advise caution on generalizing our findings outside of Java and C/C++ datasets. 

Another threat to validity is the plethora of implementation decisions we had to make for our baselines.
Each baseline discussed in section~\ref{sec:baselines} have different hyperparameters. 
These hyperparameters may change how label smoothing affect model performance.
Each model is also different in terms of parameters tuned during training (model size).
Due to hardware limitations, we do not compute the effect of label smoothing on each model using different model-intrinsic hyperparameters. Instead,
we mitigate this threat by taking the best hyperparameters listed for each baseline.
However, different hyperparameters could alter our conclusions.

%One final threat to validity is that the automatic metrics used to evaluate model performance may not be representative of human judgement.
%We mitigate this threat by using 3 different metrics: METEOR and USE+c was recently shown to better correlate with human ratings while BLEU is reported to allow direct comparison with previous work that do not report these ``newer'' metrics.

%\section{Experiment Results}
%In this section, we present our experimental findings and answer the research questions we raise in Section~\ref{sec:rqs}.
%We use RQ1 to evaluate the effect of label smoothing over a breadth of models and datasets.
%We use RQ2 to dig deeper into select models using a single dataset to evaluate the effect of the key label smoothing parameter, $\epsilon$.
%We use RQ3 to understand how label smoothing improves the performance of source code summarization models.

\section{RQ1 - Overall Performance}

This section discusses the methodology for answering RQ1 as well as our key findings.  In general, we find that label smoothing leads to improvements to several baselines.

\begin{table*}
	\small
	\centering
	\caption{Effect of label smoothing on Java q90 dataset ($\epsilon=0.1$, $N_t=10908$)}
	\vspace{-0.2cm}
	\label{tab:rq1q90}
	\begin{tabular}{p{2.2cm}|p{0.9cm}p{0.9cm}p{0.9cm}|p{0.9cm}p{0.9cm}p{0.9cm}|p{0.9cm}p{0.9cm}p{0.9cm}|p{2cm}p{2cm}}
					& \multicolumn{3}{c|}{\emph{without label smoothing}}  & \multicolumn{3}{c|}{\emph{with label smoothing}} 		&
					 \multicolumn{3}{c|}{\emph{percentage difference}}& \multicolumn{2}{c}{\emph{p-value}} \\
					 
	Model Name 		& \multicolumn{1}{c}{METEOR} 		& \multicolumn{1}{c}{USE} 	& \multicolumn{1}{c|}{BLEU} 
					& \multicolumn{1}{c}{METEOR} 		& \multicolumn{1}{c}{USE} 		& \multicolumn{1}{c|}{BLEU} 
					& \multicolumn{1}{c}{METEOR} 		& \multicolumn{1}{c}{USE} 		& \multicolumn{1}{c|}{BLEU}
					& \multicolumn{1}{c}{METEOR}   & \multicolumn{1}{c}{USE}\\ \cline{1-12} 
					
	Attendgru       & \multicolumn{1}{c}{32.82} 		& \multicolumn{1}{c}{50.21} 	& \multicolumn{1}{c|}{18.87} 			
					& \multicolumn{1}{c}{33.39} 		& \multicolumn{1}{c}{51.41} 	& \multicolumn{1}{c|}{19.45}
					& \multicolumn{1}{c}{\textcolor{white}{-}1.74\%} 		& \multicolumn{1}{c}{2.39\%} 	& \multicolumn{1}{c|}{\textcolor{white}{-}3.07\%}
					& \multicolumn{1}{r}{$<$0.01} 			& \multicolumn{1}{r}{$<$0.01} \\
	Transformer  	& \multicolumn{1}{c}{33.64} 		& \multicolumn{1}{c}{51.81} 	& \multicolumn{1}{c|}{18.99} 			
					& \multicolumn{1}{c}{33.75} 		& \multicolumn{1}{c}{52.25} 	& \multicolumn{1}{c|}{19.11} & \multicolumn{1}{c}{0.33\%} 		& \multicolumn{1}{c}{0.85\%} 	& \multicolumn{1}{c|}{0.63\%}
					& \multicolumn{1}{r}{\textcolor{white}{<}0.49} 			& \multicolumn{1}{r}{$<$0.01} \\
	AstAttendgru    & \multicolumn{1}{c}{32.81} 		& \multicolumn{1}{c}{50.20} 	& \multicolumn{1}{c|}{18.61} 			
					& \multicolumn{1}{c}{33.20} 		& \multicolumn{1}{c}{50.88} 	& \multicolumn{1}{c|}{19.12} & \multicolumn{1}{c}{1.19\%} 		& \multicolumn{1}{c}{1.35\%} 	& \multicolumn{1}{c|}{2.74\%}
					& \multicolumn{1}{r}{\textcolor{white}{<}0.02} 			& \multicolumn{1}{r}{$<$0.01} \\
	Code2Seq		& \multicolumn{1}{c}{28.86} 		& \multicolumn{1}{c}{43.99} 		& \multicolumn{1}{c|}{14.95} 			
					& \multicolumn{1}{c}{30.25} 		& \multicolumn{1}{c}{46.37} 		& \multicolumn{1}{c|}{16.19}
					 & \multicolumn{1}{c}{4.82\%} 		& \multicolumn{1}{c}{5.41\%} 	& \multicolumn{1}{c|}{8.29\%}
					& \multicolumn{1}{r}{$<$0.01} 			& \multicolumn{1}{r}{$<$0.01} \\
	CodeGNNGRU 		& \multicolumn{1}{c}{32.30} 		& \multicolumn{1}{c}{49.37} 	& \multicolumn{1}{c|}{18.00} 			
					& \multicolumn{1}{c}{33.17} 		& \multicolumn{1}{c}{50.49} 	& \multicolumn{1}{c|}{18.76}
					& \multicolumn{1}{c}{2.69\%} 		& \multicolumn{1}{c}{2.27\%} 	& \multicolumn{1}{c|}{4.22\%}
					& \multicolumn{1}{r}{$<$0.01} 			& \multicolumn{1}{r}{$<$0.01} \\
	%HAN			    & \multicolumn{1}{c}{0.00} 			& \multicolumn{1}{c}{0.00} 		& \multicolumn{1}{c|}{0.00} 			
					%& \multicolumn{1}{c}{0.00} 			& \multicolumn{1}{c}{0.00} 		& \multicolumn{1}{c|}{0.00}
					%& \multicolumn{1}{c}{0.00} 			& \multicolumn{1}{c}{0.00} \\
	AstAttendgru-fc & \multicolumn{1}{c}{33.09} 		& \multicolumn{1}{c}{50.05} 	& \multicolumn{1}{c|}{18.76} 			
					& \multicolumn{1}{c}{33.66} 		& \multicolumn{1}{c}{50.91}		& \multicolumn{1}{c|}{19.38}
					& \multicolumn{1}{c}{1.72\%} 		& \multicolumn{1}{c}{1.72\%} 	& \multicolumn{1}{c|}{3.30\%}
					& \multicolumn{1}{r}{$<$0.01} 			& \multicolumn{1}{r}{$<$0.01} \\
	\end{tabular}
\end{table*}

\begin{table*}
	%\vspace{0.2cm}
	\small
	\centering
	\caption{Effect of label smoothing on C/C++ q75 dataset ($\epsilon=0.1$, $N_t=10908$)}
	\vspace{-0.2cm}
	\label{tab:rq1q75}
	\begin{tabular}{p{2.2cm}|p{0.9cm}p{0.9cm}p{0.9cm}|p{0.9cm}p{0.9cm}p{0.9cm}|p{0.9cm}p{0.9cm}p{0.9cm}|p{2cm}p{2cm}}
		& \multicolumn{3}{c|}{\emph{without label smoothing}}  & \multicolumn{3}{c|}{\emph{with label smoothing}} 		&
		\multicolumn{3}{c|}{\emph{percentage difference}}& \multicolumn{2}{c}{\emph{p-value}} \\
		Model Name 		& \multicolumn{1}{c}{METEOR} 		& \multicolumn{1}{c}{USE} 		& \multicolumn{1}{c|}{BLEU} 
		& \multicolumn{1}{c}{METEOR} 		& \multicolumn{1}{c}{USE} 		& \multicolumn{1}{c|}{BLEU} 
		& \multicolumn{1}{c}{METEOR} 		& \multicolumn{1}{c}{USE} 		& \multicolumn{1}{c|}{BLEU}
		& \multicolumn{1}{c}{METEOR}   & \multicolumn{1}{c}{USE}\\ \cline{1-12} 
	Attendgru       & \multicolumn{1}{c}{52.38} 		& \multicolumn{1}{c}{62.25} 		& \multicolumn{1}{c|}{46.50} 			
					& \multicolumn{1}{c}{53.26} 		& \multicolumn{1}{c}{62.75} 		& \multicolumn{1}{c|}{47.41}
					& \multicolumn{1}{c}{\textcolor{white}{-}1.68\%} 		& \multicolumn{1}{c}{0.80\%} 	& \multicolumn{1}{c|}{1.96\%}
					& \multicolumn{1}{r}{$<$0.01} 		& \multicolumn{1}{r}{$<$0.01} \\
	Transformer  & \multicolumn{1}{c}{55.90} 		& \multicolumn{1}{c}{65.23} 		& \multicolumn{1}{c|}{49.30} 			
					& \multicolumn{1}{c}{56.47} 		& \multicolumn{1}{c}{65.71} 		& \multicolumn{1}{c|}{50.15}
						& \multicolumn{1}{c}{1.02\%} 		& \multicolumn{1}{c}{0.74\%} 	& \multicolumn{1}{c|}{1.72\%}
					& \multicolumn{1}{r}{$<$0.01} 		& \multicolumn{1}{r}{$<$0.01} \\
	AstAttendgru    & \multicolumn{1}{c}{48.09} 		& \multicolumn{1}{c}{57.62} 		& \multicolumn{1}{c|}{42.14} 			
					& \multicolumn{1}{c}{52.12} 		& \multicolumn{1}{c}{61.27} 		& \multicolumn{1}{c|}{46.43}
						& \multicolumn{1}{c}{8.38\%} 		& \multicolumn{1}{c}{6.33\%} 	& \multicolumn{1}{c|}{10.18\%}
					& \multicolumn{1}{r}{$<$0.01} 		& \multicolumn{1}{r}{$<$0.01} \\
	Code2seq		& \multicolumn{1}{c}{16.54} 		& \multicolumn{1}{c}{25.67} 		& \multicolumn{1}{c|}{10.20} 			
					& \multicolumn{1}{c}{17.93} 		& \multicolumn{1}{c}{28.31} 		& \multicolumn{1}{c|}{11.20}
						& \multicolumn{1}{c}{8.40\%} 		& \multicolumn{1}{c}{10.28\%} 	& \multicolumn{1}{c|}{9.80\%}
					& \multicolumn{1}{r}{$<$0.01} 		& \multicolumn{1}{r}{$<$0.01} \\
	CodeGNNGRU		& \multicolumn{1}{c}{47.79} 		& \multicolumn{1}{c}{57.17} 		& \multicolumn{1}{c|}{42.13} 			
					& \multicolumn{1}{c}{51.25} 		& \multicolumn{1}{c}{60.35} 		& \multicolumn{1}{c|}{45.63}
						& \multicolumn{1}{c}{7.24\%} 		& \multicolumn{1}{c}{5.56\%} 	& \multicolumn{1}{c|}{8.31\%}
					& \multicolumn{1}{r}{$<$0.01} 		& \multicolumn{1}{r}{$<$0.01} \\
%	HAN        & \multicolumn{1}{c}{0.00} 		& \multicolumn{1}{c}{0.00} 		& \multicolumn{1}{c|}{0.00} 			
	%				& \multicolumn{1}{c}{0.00} 		& \multicolumn{1}{c}{0.00} 		& \multicolumn{1}{c|}{0.00}
	%				& \multicolumn{1}{c}{0.00} 		& \multicolumn{1}{c}{0.00} \\
	AstAttendgru-fc & \multicolumn{1}{c}{46.68} 		& \multicolumn{1}{c}{56.61} 		& \multicolumn{1}{c|}{40.79} 			
					& \multicolumn{1}{c}{50.66} 		& \multicolumn{1}{c}{60.32} 		& \multicolumn{1}{c|}{44.62}
						& \multicolumn{1}{c}{8.53\%} 		& \multicolumn{1}{c}{6.55\%} 	& \multicolumn{1}{c|}{9.39\%}
					& \multicolumn{1}{r}{$<$0.01} 		& \multicolumn{1}{r}{$<$0.01} \\
	\end{tabular}
\end{table*}

\begin{table*}
	\small
	%\vspace{0.2cm}
	\centering
	\caption{Effect of label smoothing on full Java dataset ($\epsilon=0.1$, $N_t=10908$)}
	\vspace{-0.2cm}
	\label{tab:rq1full}
	\begin{tabular}{p{2.2cm}|p{0.9cm}p{0.9cm}p{0.9cm}|p{0.9cm}p{0.9cm}p{0.9cm}|p{0.9cm}p{0.9cm}p{0.9cm}|p{2cm}p{2cm}}
		& \multicolumn{3}{c|}{\emph{without label smoothing}}  & \multicolumn{3}{c|}{\emph{with label smoothing}} 		&
		\multicolumn{3}{c|}{\emph{percentage difference}}& \multicolumn{2}{c}{\emph{p-value}} \\
		Model Name 		& \multicolumn{1}{c}{METEOR} 		& \multicolumn{1}{c}{USE} 		& \multicolumn{1}{c|}{BLEU} 
		& \multicolumn{1}{c}{METEOR} 		& \multicolumn{1}{c}{USE} 		& \multicolumn{1}{c|}{BLEU} 
		& \multicolumn{1}{c}{METEOR} 		& \multicolumn{1}{c}{USE} 		& \multicolumn{1}{c|}{BLEU}
		& \multicolumn{1}{c}{METEOR}   & \multicolumn{1}{c}{USE}\\ \cline{1-12} 
		Attendgru       & \multicolumn{1}{c}{35.40} 		& \multicolumn{1}{c}{52.84} 		& \multicolumn{1}{c|}{18.44} 			
		& \multicolumn{1}{c}{35.21} 		& \multicolumn{1}{c}{53.08} 		& \multicolumn{1}{c|}{18.35}
		& \multicolumn{1}{c}{-0.54\%} 		& \multicolumn{1}{c}{0.45\%} 	& \multicolumn{1}{c|}{-0.49\%}
		& \multicolumn{1}{r}{$<$0.01} 		& \multicolumn{1}{r}{$<$0.01}\\
		Transformer  & \multicolumn{1}{c}{36.44} 		& \multicolumn{1}{c}{53.95} 		& \multicolumn{1}{c|}{19.12} 			
		& \multicolumn{1}{c}{36.87} 		& \multicolumn{1}{c}{54.58} 		& \multicolumn{1}{c|}{19.36}
		& \multicolumn{1}{c}{1.18\%} 		& \multicolumn{1}{c}{1.17\%} 	& \multicolumn{1}{c|}{1.26\%}
		& \multicolumn{1}{r}{$<$0.01} 		& \multicolumn{1}{r}{$<$0.01}\\
		AstAttendgru    & \multicolumn{1}{c}{35.72} 		& \multicolumn{1}{c}{53.03} 		& \multicolumn{1}{c|}{18.53} 			
		& \multicolumn{1}{c}{35.90} 		& \multicolumn{1}{c}{53.44} 		& \multicolumn{1}{c|}{18.77}
		& \multicolumn{1}{c}{0.50\%} 		& \multicolumn{1}{c}{0.77\%} 	& \multicolumn{1}{c|}{1.30\%}
		& \multicolumn{1}{r}{$<$0.01} 		& \multicolumn{1}{r}{$<$0.01}\\
		Code2seq		& \multicolumn{1}{c}{35.38} 		& \multicolumn{1}{c}{52.96} 		& \multicolumn{1}{c|}{18.32} 			
		& \multicolumn{1}{c}{35.87} 		& \multicolumn{1}{c}{53.45} 		& \multicolumn{1}{c|}{18.62}
		& \multicolumn{1}{c}{1.38\%} 		& \multicolumn{1}{c}{0.93\%} 	& \multicolumn{1}{c|}{1.64\%}
		& \multicolumn{1}{r}{$<$0.01} 		& \multicolumn{1}{r}{ $<$0.01}\\
		CodeGNNGRU		& \multicolumn{1}{c}{36.01} 		& \multicolumn{1}{c}{53.57} 		& \multicolumn{1}{c|}{18.91} 			
		& \multicolumn{1}{c}{36.09} 		& \multicolumn{1}{c}{53.82} 		& \multicolumn{1}{c|}{18.89}
		& \multicolumn{1}{c}{0.22\%} 		& \multicolumn{1}{c}{0.47\%} 	& \multicolumn{1}{c|}{-0.11\%}
		& \multicolumn{1}{r}{\textcolor{white}{<}0.20} 		& \multicolumn{1}{r}{$<$0.01}\\
		AstAttendgru-fc & \multicolumn{1}{c}{35.93} 		& \multicolumn{1}{c}{53.39} 		& \multicolumn{1}{c|}{19.14} 			
		& \multicolumn{1}{c}{36.37} 		& \multicolumn{1}{c}{53.86} 		& \multicolumn{1}{c|}{19.61}
		& \multicolumn{1}{c}{1.22\%} 		& \multicolumn{1}{c}{0.88\%} 	& \multicolumn{1}{c|}{2.46\%}
		& \multicolumn{1}{r}{$<$0.01} 		& \multicolumn{1}{r}{$<$0.01}\\
	\end{tabular}
\end{table*}

\subsubsection*{Methodology}

To answer RQ1, we follow the established methodology that have become standard in source code summarization and NMT tasks in NLP~\cite{leclair2019neural, vaswani2017attention}.
We use two standard datasets in two common programming languages.
We perform further filtration on the dataset by taking the top 10\% largest methods from the Java dataset (Java-q90) and the top-25\% largest methods from the C/C++ dataset (C/C++-q75).
We do this to find functions that have a large number of statements, which is more representative of real world use-case scenario.
We also train the models on the full 1.9m Java dataset to evaluate model performance on a full dataset to make commensurable comparison with related work.
For all three dataset instances, we train each baseline twice: once with and again without label smoothing.
We keep the value of $\epsilon$ constant (0.1) for all models.
For each architecture, we train both instances (with and without label smoothing) for 10 epochs and choose the model with the highest validation accuracy score for our comparison (standard practice in related work).
We then evaluate the performance of these models using automated evaluation techniques discussed in Section~\ref{sec:metrics}.
Roy~\emph{et al.} at FSE'21 recommended to perform a paired t-test between baseline model predictions and new model predictions for different sentence-level metrics.
Therefore, we perform a paired t-test to identify if the performance difference between two comparing model predictions is statistically significant.
Note that we do not perform a t-test on BLEU score because BLEU is a corpus level metric.
%The test is paired because the prediction files are related and one-sided because we are only interested if the performance of a model improves with label smoothing.

\subsubsection*{Key Findings}

Our key finding in answering RQ1 is that adding label smoothing as a regularizer improves model performance in most cases.
All models show performance improvement in terms of USE for Java-q90, C/C++-q75 and full Java dataset.
All models using the Java-q90 and C/C++-q75 datasets and most models using the full Java dataset also report performance improvement in term of METEOR and BLEU.
Furthermore, most of the performance improvements are statistically significant (we choose $\alpha=0.05$).
This overall trend of performance improvement is depicted in Figure~\ref{fig:rq1}.
The dark blue bar shows the METEOR scores for models without label smoothing.
The orange bar on top shows the increase in the METEOR score after we add label smoothing.
As the figures show, there is an increase in METEOR score in almost all cases.

Table~\ref{tab:rq1q90} shows the effect of label smoothing on the overall performance of each baseline for the Java-q90 dataset.
We find that all models report higher evaluation metric score with the addition of label smoothing. 
However, Code2seq improves the most with 4.8\% improvement on METEOR, 5.4\% improvement on USE and 8.3\% improvement on BLEU.
Codegnngru also shows significant improvement with 2.7\% improvement on METEOR, 2.3\% improvement on USE and 4.2\% improvement on BLEU.
Attendgru and ast-attendgru-fc also show similar improvement with 1.7\% improvement for both on METEOR, 2.4\% and 1.7\% improvement on USE and 3.1\% and 3.3\% improvement on BLEU respectively.
Each of these models vary with respect to encoder input.
Each model has different inputs as well as different data structure for common inputs.
The higher scores are encouraging as we observe statistically significant performance improvement across different model architectures.
Transformers, however, show the least overall improvement in performance with an increment of less than 1\% for all metrics, although the 0.85\% increase in USE score is statistically significant.
One possible explanation for this result is that Transformers have built-in dropout layers after each multi-head attention layer.  These dropout layers likely already improve regularization.

Table~\ref{tab:rq1q75} shows the effect of label smoothing on the overall performance of each baseline for the C/C++-q75 dataset.
For this dataset, we notice a large improvement in performance for ast-attendgru, code2seq, codegnngru and ast-attendgru-fc baselines.
These performance increase range from 7.2\%-8.5\% for METEOR, 5.6\%-10.3\% for USE and 8.3\%-10.2\% for BLEU.
It is interesting to note that attendgru does not show a large percentage increase in performance ($<$2\% across all metrics) for this dataset, compared to the aforementioned models.
Furthermore, similar to the Java-q90 dataset, transformers also show little improvement (about 1-2\% across all metrics).
We expect this result for transformers due to the built-in dropout regularizer in the transformer architecture.
However, the performance increase for both attendgru and transformers is still statistically significant with p-values $<$0.01 for both METEOR and USE.

Table~\ref{tab:rq1full} shows the effect of label smoothing on the overall performance of each baseline for the full Java dataset of 1.9m methods.
Again we see a trend of performance improvement in most cases.
However, while the increase in metric score is significant in all but two configurations, we do notice that the percentage increment is not as high as for the Java-q90 and C/C++-q75 dataset ($<$2\% across all metrics for all models).
We attribute this to the fact that higher training examples act as a regularizer in and of itself.
For attendgru, we notice a statistically significant decrease in METEOR score but a statistically significant increase in USE score.
We also notice a small decrease in BLEU score.
Since all three metrics do not point in one way or another, we do not draw any conclusion for this setup.
For codegnngru, we notice a 0.11\% decrease in BLEU score, a 0.22\% increase in METEOR score which is not statistically significant but a 0.47\% increase in USE score which is statistically significant.
Since the two metric scores that show performance improvement are also correlated best with human evaluation, we conclude that label smoothing positively affects codegnngru.
Once again, code2seq shows the highest performance increase for the Java full dataset.
One likely explanation for why code2seq achieves the highest performance increase is that it is the largest model among the baselines.  However, its parameters do require more time to train than others.
Like for other models, label smoothing appears to help code2seq generalize.% because of this.

%\newpage
\begin{table}[!t]
	\small
	\caption{\small{Scores of attendgru for different $\epsilon$. Output vocabulary size for the top table is 10k and for the bottom table is 44k.}}
	\vspace{-0.1cm}
	\begin{tabular}{p{0.6cm}|p{0.8cm}p{0.8cm}p{0.8cm}|p{0.5cm}p{0.5cm}}
		& \multicolumn{3}{c|}{\emph{metric scores}} & \multicolumn{2}{c}{\emph{t-stat, p-value}} \\
		$\epsilon$ 	& \multicolumn{1}{c}{\small{METEOR}} 		& \multicolumn{1}{c}{\small{USE}} 	& \multicolumn{1}{c|}{\small{BLEU}}
		& \multicolumn{1}{c}{\small{METEOR}}  & \multicolumn{1}{c}{\small{USE}} \\  \cline{1-6}
		0       & \multicolumn{1}{c}{32.82} & \multicolumn{1}{c}{50.21} & \multicolumn{1}{c|}{18.87} 
		& \multicolumn{1}{c}{-} 	& \multicolumn{1}{c}{-} \\
		%& \multicolumn{1}{c}{\emph{t-stat, p-value}} 	& \multicolumn{1}{c}{\emph{t-stat, p-value}} \\
		0.001       & \multicolumn{1}{c}{32.79} & \multicolumn{1}{c}{50.22} & \multicolumn{1}{c|}{18.85} 			
		& \multicolumn{1}{r}{-0.26,\textcolor{white}{$<$}0.80} 	& \multicolumn{1}{r}{\textcolor{white}{$<$}0.13,\textcolor{white}{$<$}0.90}  \\
		0.003       & \multicolumn{1}{c}{33.01} & \multicolumn{1}{c}{50.87} & \multicolumn{1}{c|}{18.82} 			
		& \multicolumn{1}{r}{1.34,\textcolor{white}{$<$}0.18} 	& \multicolumn{1}{r}{4.60,$<$0.01} \\
		0.007       & \multicolumn{1}{c}{32.99} & \multicolumn{1}{c}{50.36} & \multicolumn{1}{c|}{18.99} 			
		& \multicolumn{1}{r}{1.32,\textcolor{white}{$<$}0.19} 	& \multicolumn{1}{r}{1.13,\textcolor{white}{$<$}0.26} \\
		0.02       	& \multicolumn{1}{c}{32.97} & \multicolumn{1}{c}{50.36} & \multicolumn{1}{c|}{19.01} 			
		& \multicolumn{1}{r}{1.07,\textcolor{white}{$<$}0.28} 	& \multicolumn{1}{r}{0.99,\textcolor{white}{$<$}0.32} \\
		0.05	    & \multicolumn{1}{c}{33.30} & \multicolumn{1}{c}{51.00} & \multicolumn{1}{c|}{19.30} 			
		& \multicolumn{1}{r}{3.18,$<$0.01} 	& \multicolumn{1}{r}{5.19,$<$0.01} \\
		0.10       	& \multicolumn{1}{c}{33.39} & \multicolumn{1}{c}{51.41} & \multicolumn{1}{c|}{19.45} 			
		& \multicolumn{1}{r}{3.76,$<$0.01} 	& \multicolumn{1}{r}{7.84,$<$0.01} \\
		0.25       	& \multicolumn{1}{c}{33.29} & \multicolumn{1}{c}{51.05} & \multicolumn{1}{c|}{19.32} 			
		& \multicolumn{1}{r}{2.93,$<$0.01} 	& \multicolumn{1}{r}{5.12,$<$0.01} \\
		0.40        & \multicolumn{1}{c}{33.34} & \multicolumn{1}{c}{50.98} & \multicolumn{1}{c|}{19.32} 			
		& \multicolumn{1}{r}{3.25,$<$0.01} 	& \multicolumn{1}{r}{4.63,$<$0.01}
	\end{tabular}
	\vspace{0.2cm}
	
	\begin{tabular}{p{0.6cm}|p{1.0cm}p{1.0cm}p{1.0cm}|p{1.5cm}p{1.5cm}}
		& \multicolumn{3}{c|}{\emph{metric scores}} & \multicolumn{2}{c}{\emph{t-stat, p-value}} \\
		$\epsilon$ 	& \multicolumn{1}{c}{\small{METEOR}} 		& \multicolumn{1}{c}{\small{USE}} 	& \multicolumn{1}{c|}{\small{BLEU}}
		& \multicolumn{1}{c}{\small{METEOR}}  & \multicolumn{1}{c}{\small{USE}} \\  \cline{1-6}
		0       & \multicolumn{1}{c}{32.94} 	& \multicolumn{1}{c}{50.30} 		& \multicolumn{1}{c|}{18.94} 			
		& \multicolumn{1}{c}{-} 	& \multicolumn{1}{c}{-} \\
		0.001       & \multicolumn{1}{c}{32.94} 	& \multicolumn{1}{c}{50.23} 		& \multicolumn{1}{c|}{18.89} 			
		& \multicolumn{1}{r}{-0.07,\textcolor{white}{$<$}0.94} 	& \multicolumn{1}{r}{-0.79,\textcolor{white}{$<$}0.43} \\
		0.003       & \multicolumn{1}{c}{32.79} 	& \multicolumn{1}{c}{50.10} 		& \multicolumn{1}{c|}{18.95} 			
		& \multicolumn{1}{r}{-1.24,\textcolor{white}{$<$}0.21} 	& \multicolumn{1}{r}{-1.66,\textcolor{white}{$<$}0.10} \\
		0.007       & \multicolumn{1}{c}{32.71} 	& \multicolumn{1}{c}{50.11} 		& \multicolumn{1}{c|}{18.86} 			
		& \multicolumn{1}{r}{-1.70,\textcolor{white}{$<$}0.09} 	& \multicolumn{1}{r}{-1.42,\textcolor{white}{$<$}0.16} \\
		0.02       	& \multicolumn{1}{c}{32.85} 	& \multicolumn{1}{c}{50.60} 		& \multicolumn{1}{c|}{18.90} 			
		& \multicolumn{1}{r}{-0.63,\textcolor{white}{$<$}0.53} 	& \multicolumn{1}{r}{2.06,\textcolor{white}{$<$}0.04} \\
		0.05	    & \multicolumn{1}{c}{32.95} 	& \multicolumn{1}{c}{50.86} 		& \multicolumn{1}{c|}{19.02} 			
		& \multicolumn{1}{r}{0.07,\textcolor{white}{$<$}0.95} 	& \multicolumn{1}{r}{3.66,$<$0.01} \\
		0.10       	& \multicolumn{1}{c}{33.20} 	& \multicolumn{1}{c}{50.97} 		& \multicolumn{1}{c|}{19.05} 			
		& \multicolumn{1}{r}{1.83,\textcolor{white}{$<$}0.07} 	& \multicolumn{1}{r}{4.37,$<$0.01} \\
		0.25       	& \multicolumn{1}{c}{33.18} 	& \multicolumn{1}{c}{50.88} 		& \multicolumn{1}{c|}{19.05} 			
		& \multicolumn{1}{r}{1.45,\textcolor{white}{$<$}0.15} 	& \multicolumn{1}{r}{3.55,$<$0.01} \\
		0.40       & \multicolumn{1}{c}{33.16} 	& \multicolumn{1}{c}{50.89} 		& \multicolumn{1}{c|}{19.22} 			
		& \multicolumn{1}{r}{1.32,\textcolor{white}{$<$}0.19} 	& \multicolumn{1}{r}{3.64,$<$0.01}
	\end{tabular}
	\vspace{-0.2cm}
	\label{tab:rq2attendgru}
\end{table}

\section{RQ2 - Hyperparameter Tuning}

\subsubsection*{Methodology}
\label{sec:rq2_method}

To answer RQ2, we choose four out of the six models used in RQ1 to study in greater detail: \texttt{attendgru}, \texttt{transformer}, \texttt{codegnngru} and \texttt{ast-attendgru-fc}.
We use these models because they each represent a different family of source code summarizaton models: \texttt{attendgru} from simple seq2seq family, \texttt{transformer} from self-attention architecture family, \texttt{codegnngru} from code + AST represented as GNN in a seq2seq architecture family and \texttt{ast-attendgru-fc} from code + flat AST + contextual information in a seq2seq model family.
For each model we vary the value of $\epsilon$; we start with 0.001 and increase the value by a factor of $e^n$ for n=1 to 6.
Along the way, we also include 0.25 (~0.001$\times e^{5.5}$) for a more granular look as $\epsilon$ increases exponentially for large values of n.
We run each model for these 8 different configurations on the Java-q90 dataset.
We only choose one dataset for RQ2 because we want to eliminate any experimental variables that may be introduced by the dataset, but also due to resource constraints.
We choose the Java-q90 dataset because it is part of a peer-vetted and widely used source code summarization dataset.
To understand how output vocabulary size affects model performance with label smoothing, we then change the size of output vocabulary more than four-fold (from 10k to 44k) and run each model again.
We train each model configuration for eight epochs and choose the model with highest validation accuracy.
Similar to RQ1, we evaluate the performance of these models using automated evaluation techniques discussed in Section~\ref{sec:metrics}.
Additionally we perform a paired t-test between each configuration with label smoothing and the corresponding prediction without label smoothing to identify the statistical significance for METEOR and USE. % as recommended by related work.
Notice again that we do not perform a t-test on BLEU score because BLEU is a corpus level metric.

\begin{table}[!t]
	\small
	\setlength\tabcolsep{5pt}
	\caption{\small{Scores of transformer for different $\epsilon$. Output vocabulary size for the top table is 10k and for the bottom table is 44k.}}
	\vspace{-0.1cm}
	\begin{tabular}{p{0.6cm}|p{0.6cm}p{0.6cm}p{0.6cm}|p{1.5cm}p{1.5cm}}
		& \multicolumn{3}{c|}{\emph{metric scores}} & \multicolumn{2}{c}{\emph{t-stat, p-value}} \\
		$\epsilon$ 	& \multicolumn{1}{c}{\small{METEOR}} 		& \multicolumn{1}{c}{\small{USE}} 	& \multicolumn{1}{c|}{\small{BLEU}}
		& \multicolumn{1}{c}{\small{METEOR}}  & \multicolumn{1}{c}{\small{USE}} \\  \cline{1-6}
		0       & \multicolumn{1}{c}{33.64} & \multicolumn{1}{c}{51.81} & \multicolumn{1}{c|}{18.99} 			
		& \multicolumn{1}{c}{-} 	& \multicolumn{1}{c}{-} \\
		0.001       & \multicolumn{1}{c}{33.53} & \multicolumn{1}{c}{51.46} & \multicolumn{1}{c|}{19.18} 			
		& \multicolumn{1}{r}{-0.83, \textcolor{white}{$<$}0.41} 	& \multicolumn{1}{r}{-2.71, $<$0.01} \\
		0.003       & \multicolumn{1}{c}{33.27} & \multicolumn{1}{c}{51.36} & \multicolumn{1}{c|}{19.06} 			
		& \multicolumn{1}{r}{-2.53, \textcolor{white}{$<$}0.01} 	& \multicolumn{1}{r}{-3.18, $<$0.01} \\
		0.007       & \multicolumn{1}{c}{33.77} & \multicolumn{1}{c}{51.49} & \multicolumn{1}{c|}{19.13} 			
		& \multicolumn{1}{r}{0.81, \textcolor{white}{$<$}0.42} 	& \multicolumn{1}{r}{-2.05, \textcolor{white}{$<$}0.04} \\
		0.02       	& \multicolumn{1}{c}{33.84} & \multicolumn{1}{c}{52.57} & \multicolumn{1}{c|}{19.08} 			
		& \multicolumn{1}{r}{1.22, \textcolor{white}{$<$}0.22} 	& \multicolumn{1}{r}{4.76, $<$0.01} \\
		0.05	    & \multicolumn{1}{c}{33.65} & \multicolumn{1}{c}{51.64} & \multicolumn{1}{c|}{19.11} 			
		& \multicolumn{1}{r}{0.07, \textcolor{white}{$<$}0.95} 	& \multicolumn{1}{r}{-1.14, \textcolor{white}{$<$}0.26} \\
		0.10       	& \multicolumn{1}{c}{33.75} & \multicolumn{1}{c}{52.25} & \multicolumn{1}{c|}{19.11} 			
		& \multicolumn{1}{r}{0.69, \textcolor{white}{$<$}0.49} 	& \multicolumn{1}{r}{2.72, $<$0.01} \\
		0.25       	& \multicolumn{1}{c}{33.72} & \multicolumn{1}{c}{52.27} & \multicolumn{1}{c|}{18.97} 			
		& \multicolumn{1}{r}{0.50, \textcolor{white}{$<$}0.62} 	& \multicolumn{1}{r}{2.96, $<$0.01} \\
		0.40        & \multicolumn{1}{c}{33.84} & \multicolumn{1}{c}{52.10} & \multicolumn{1}{c|}{19.19} 			
		& \multicolumn{1}{r}{1.21, \textcolor{white}{$<$}0.23} 	& \multicolumn{1}{r}{1.73, \textcolor{white}{$<$}0.08}
	\end{tabular}
	\vspace{0.2cm}
	
	\begin{tabular}{p{0.6cm}|p{0.6cm}p{0.6cm}p{0.6cm}|p{1.5cm}p{1.5cm}}
		& \multicolumn{3}{c|}{\emph{metric scores}} & \multicolumn{2}{c}{\emph{t-stat, p-value}} \\
		$\epsilon$ 	& \multicolumn{1}{c}{\small{METEOR}} 		& \multicolumn{1}{c}{\small{USE}} 	& \multicolumn{1}{c|}{\small{BLEU}}
		& \multicolumn{1}{c}{\small{METEOR}}  & \multicolumn{1}{c}{\small{USE}} \\  \cline{1-6}
		0       & \multicolumn{1}{c}{33.28} 	& \multicolumn{1}{c}{51.55} 		& \multicolumn{1}{c|}{18.73} 			
		& \multicolumn{1}{c}{-} 	& \multicolumn{1}{c}{-} \\
		0.001       & \multicolumn{1}{c}{33.38} 	& \multicolumn{1}{c}{50.95} 		& \multicolumn{1}{c|}{18.85} 			
		& \multicolumn{1}{r}{0.62, \textcolor{white}{$<$}0.54} 	& \multicolumn{1}{r}{-3.67, $<$0.01} \\
		0.003       & \multicolumn{1}{c}{33.09} 	& \multicolumn{1}{c}{51.23} 		& \multicolumn{1}{c|}{18.77} 			
		& \multicolumn{1}{r}{-1.21, \textcolor{white}{$<$}0.23} 	& \multicolumn{1}{r}{-2.11, \textcolor{white}{$<$}0.04} \\
		0.007       & \multicolumn{1}{c}{33.30} 	& \multicolumn{1}{c}{51.55} 		& \multicolumn{1}{c|}{18.75} 			
		& \multicolumn{1}{r}{-0.83, \textcolor{white}{$<$}0.41} 	& \multicolumn{1}{r}{0.01, \textcolor{white}{$<$}0.99} \\
		0.02       	& \multicolumn{1}{c}{33.30} 	& \multicolumn{1}{c}{51.82} 		& \multicolumn{1}{c|}{18.85} 			
		& \multicolumn{1}{r}{0.13, \textcolor{white}{$<$}0.90} 	& \multicolumn{1}{r}{1.71, \textcolor{white}{$<$}0.09} \\
		0.05	    & \multicolumn{1}{c}{33.32} 	& \multicolumn{1}{c}{51.74} 		& \multicolumn{1}{c|}{18.77} 			
		& \multicolumn{1}{r}{0.27, \textcolor{white}{$<$}0.79} 	& \multicolumn{1}{r}{1.21, \textcolor{white}{$<$}0.22} \\
		0.10       	& \multicolumn{1}{c}{33.77} 	& \multicolumn{1}{c}{52.17} 		& \multicolumn{1}{c|}{18.94} 			
		& \multicolumn{1}{r}{3.04, $<$0.01} 	& \multicolumn{1}{r}{3.89, $<$0.01} \\
		0.25       	& \multicolumn{1}{c}{33.84} 	& \multicolumn{1}{c}{52.51} 		& \multicolumn{1}{c|}{19.22} 			
		& \multicolumn{1}{r}{3.40, $<$0.01} 	& \multicolumn{1}{r}{5.95, $<$0.01} \\
		0.40       & \multicolumn{1}{c}{34.23} 	& \multicolumn{1}{c}{52.27} 		& \multicolumn{1}{c|}{19.56} 			
		& \multicolumn{1}{r}{5.57, $<$0.01} 	& \multicolumn{1}{r}{4.33, $<$0.01}
	\end{tabular}
	\vspace{-0.4cm}
	\label{tab:rq2transformer}
\end{table}

\subsubsection*{Key Findings}

Our key finding in answering RQ2 is that higher values of $\epsilon$ generally lead to higher model performance, although this increase in performance is less pronounced for $\epsilon>$ 0.1.
Furthermore, this trend is unaffected as we increase the vocabulary size from 10k to 44k.
Therefore, the value of the smoothed probability per token ($\frac{\epsilon}{N_t-1}$) does not affect model performance.
While one must decide the best hyperparameters for oneself based on prevailing experimental conditions, our recommendation is to set $\epsilon$=0.1 for initial evaluation.
%Additional evaluation of these models for higher values of $\epsilon$ might then be worthwhile to squeeze out the highest metric score.
Caution is advised for high values of $\epsilon$ as it forces models to predict more common words and eliminate rare/unique words.
We discuss this issue in greater detail in RQ3.

Table~\ref{tab:rq2attendgru} shows the effect of increasing the label smoothing factor, $\epsilon$ for attendgru model.
For this architecture, we notice an increase in model performance on all metrics as we increase the value of $\epsilon$ from 0.001 to 0.1.
METEOR, USE and BLEU scores all achieve the highest score for this configuration of $\epsilon$=0.1.
%While performance dips slightly as we further increase $\epsilon$ to 0.25 and 0.4, it is still significantly higher than the model instance with no label smoothing ($\epsilon$=0).
As we increase the output vocabulary size from 10k to 44k, we notice that both METEOR and USE scores decrease slightly for small values of $\epsilon$ but then they increase, reaching highest METEOR and USE score for $\epsilon$=0.1.

\begin{table}[!t]
	\small
	\setlength\tabcolsep{5pt}
	\caption{\small{Scores of codegnngru for different $\epsilon$. Output vocabulary size for the top table is 10k and for the bottom table is 44k.}}
	\vspace{-0.1cm}
	\begin{tabular}{p{0.6cm}|p{1.0cm}p{1.0cm}p{1.0cm}|p{1.5cm}p{1.5cm}}
		& \multicolumn{3}{c|}{\emph{metric scores}} & \multicolumn{2}{c}{\emph{t-stat, p-value}} \\
		$\epsilon$ 	& \multicolumn{1}{c}{\small{METEOR}} 		& \multicolumn{1}{c}{\small{USE}} 	& \multicolumn{1}{c|}{\small{BLEU}}
		& \multicolumn{1}{c}{\small{METEOR}}  & \multicolumn{1}{c}{\small{USE}} \\  \cline{1-6}
		0	        & \multicolumn{1}{c}{32.30} & \multicolumn{1}{c}{49.37} & \multicolumn{1}{c|}{18.00} 			
		& \multicolumn{1}{c}{-} 	& \multicolumn{1}{c}{-} \\
		0.001       & \multicolumn{1}{c}{32.53} & \multicolumn{1}{c}{49.94} & \multicolumn{1}{c|}{18.22} 			
		& \multicolumn{1}{r}{1.55, \textcolor{white}{$<$}0.12} 	& \multicolumn{1}{r}{3.64, $<$0.01} \\
		0.003       & \multicolumn{1}{c}{32.46} & \multicolumn{1}{c}{49.76} & \multicolumn{1}{c|}{18.24} 			
		& \multicolumn{1}{r}{0.98, \textcolor{white}{$<$}0.33} 	& \multicolumn{1}{r}{2.30, \textcolor{white}{$<$}0.02} \\
		0.007       & \multicolumn{1}{c}{32.18} & \multicolumn{1}{c}{49.69} & \multicolumn{1}{c|}{17.96} 			
		& \multicolumn{1}{r}{-0.79, \textcolor{white}{$<$}0.43} 	& \multicolumn{1}{r}{2.06, \textcolor{white}{$<$}0.04} \\
		0.02       	& \multicolumn{1}{c}{32.53} & \multicolumn{1}{c}{49.88} & \multicolumn{1}{c|}{18.42} 			
		& \multicolumn{1}{r}{1.40, \textcolor{white}{$<$}0.16} 	& \multicolumn{1}{r}{3.02, $<$0.01} \\
		0.05	    & \multicolumn{1}{c}{32.65} & \multicolumn{1}{c}{49.92} & \multicolumn{1}{c|}{18.59} 			
		& \multicolumn{1}{r}{2.16, \textcolor{white}{$<$}0.03} 	& \multicolumn{1}{r}{3.22, $<$0.01} \\
		0.10       	& \multicolumn{1}{c}{33.17} & \multicolumn{1}{c}{50.49} & \multicolumn{1}{c|}{18.76} 			
		& \multicolumn{1}{r}{5.46, $<$0.01} 	& \multicolumn{1}{r}{6.82, $<$0.01} \\
		0.25       	& \multicolumn{1}{c}{33.36} & \multicolumn{1}{c}{51.08} & \multicolumn{1}{c|}{18.97} 			
		& \multicolumn{1}{r}{6.65, \textcolor{white}{$<$}0.01} 	& \multicolumn{1}{r}{10.19, $<$0.01} \\
		0.40        & \multicolumn{1}{c}{33.37} & \multicolumn{1}{c}{51.35} & \multicolumn{1}{c|}{19.00} 			
		& \multicolumn{1}{r}{6.51, $<$0.01} 	& \multicolumn{1}{r}{11.79, $<$0.01}
	\end{tabular}
	\vspace{0.2cm}
	
	\begin{tabular}{p{0.6cm}|p{0.6cm}p{0.6cm}p{0.6cm}|p{1.5cm}p{1.5cm}}
		& \multicolumn{3}{c|}{\emph{metric scores}} & \multicolumn{2}{c}{\emph{t-stat, p-value}} \\
		$\epsilon$ 	& \multicolumn{1}{c}{\small{METEOR}} 		& \multicolumn{1}{c}{\small{USE}} 	& \multicolumn{1}{c|}{\small{BLEU}}
		& \multicolumn{1}{c}{\small{METEOR}}  & \multicolumn{1}{c}{\small{USE}} \\  \cline{1-6}
		0	        & \multicolumn{1}{c}{32.26} 	& \multicolumn{1}{c}{49.33} 		& \multicolumn{1}{c|}{18.20} 			
		& \multicolumn{1}{c}{-} 	& \multicolumn{1}{c}{-} \\
		0.001       & \multicolumn{1}{c}{32.56} 	& \multicolumn{1}{c}{49.85} 		& \multicolumn{1}{c|}{18.22} 			
		& \multicolumn{1}{r}{1.87, \textcolor{white}{$<$}0.06} 	& \multicolumn{1}{r}{3.31, $<$0.01} \\
		0.003       & \multicolumn{1}{c}{32.25} 	& \multicolumn{1}{c}{49.62} 		& \multicolumn{1}{c|}{18.19} 			
		& \multicolumn{1}{r}{-0.07, \textcolor{white}{$<$}0.94} 	& \multicolumn{1}{r}{1.77, \textcolor{white}{$<$}0.08} \\
		0.007       & \multicolumn{1}{c}{32.75} 	& \multicolumn{1}{c}{49.68} 		& \multicolumn{1}{c|}{18.57} 			
		& \multicolumn{1}{r}{3.02, $<$0.01} 	& \multicolumn{1}{r}{2.10, \textcolor{white}{$<$}0.04} \\
		0.02       	& \multicolumn{1}{c}{32.36} 	& \multicolumn{1}{c}{49.76} 		& \multicolumn{1}{c|}{18.34} 			
		& \multicolumn{1}{r}{0.62, \textcolor{white}{$<$}0.53} 	& \multicolumn{1}{r}{2.68, $<$0.01} \\
		0.05	    & \multicolumn{1}{c}{32.86} 	& \multicolumn{1}{c}{50.31} 		& \multicolumn{1}{c|}{18.53} 			
		& \multicolumn{1}{r}{3.65, $<$0.01} 	& \multicolumn{1}{r}{5.98, $<$0.01} \\
		0.10       	& \multicolumn{1}{c}{32.71} 	& \multicolumn{1}{c}{49.97} 		& \multicolumn{1}{c|}{18.48} 			
		& \multicolumn{1}{r}{2.64, $<$0.01} 	& \multicolumn{1}{r}{3.77, $<$0.01} \\
		0.25       	& \multicolumn{1}{c}{32.68} 	& \multicolumn{1}{c}{49.92} 		& \multicolumn{1}{c|}{18.53} 			
		& \multicolumn{1}{r}{2.51, \textcolor{white}{$<$}0.01} 	& \multicolumn{1}{r}{3.53, $<$0.01} \\
		0.40       & \multicolumn{1}{c}{32.88} 	& \multicolumn{1}{c}{50.50} 		& \multicolumn{1}{c|}{18.72} 			
		& \multicolumn{1}{r}{3.72, $<$0.01} 	& \multicolumn{1}{r}{7.11, $<$0.01}
	\end{tabular}
	\vspace{-0.4cm}
	\label{tab:rq2codegnngru}
\end{table}

Table~\ref{tab:rq2transformer} shows the effect of increasing $\epsilon$ for transformers.
We initially see a decrease in metric scores for small values of $\epsilon$.
As we increase $\epsilon$, the metric scores start to increase.
Interestingly, for the 44k output vocabulary, we see a substantially significant increase in metric scores for all values of $\epsilon >$ 0.05.
We attribute this to the fact that higher output vocabulary size increases the amount of rare words in the prediction. 
Label smoothing helps models generalize by focusing on more commonly occuring words.
%This finding is further validated in RQ3.

Table~\ref{tab:rq2codegnngru} shows the effect of increasing $\epsilon$ for codegnngru.
For the 10k output vocabulary set, we notice a pattern similar to attendgru and transformers.
%The metrics scores change slightly for small values of $\epsilon$.
While USE shows a statistically significant improvement for all cases of $\epsilon$, METEOR only starts demonstrating a statistically significant improvement for all values of $\epsilon >$ 0.05.
For the 44k output vocabulary set, we see an improvement across the board, that is explained by the model eliminating noise, introduced by the large output vocabulary, set using label smoothing.

Table~\ref{tab:rq2astattendgrufc} shows the effect of increasing $\epsilon$ for astattendgru-fc.
For the 10k output vocabulary set, we see a statistically significant decrease in the model performance for $\epsilon$=0.001.
For $\epsilon$=0.003, we see a slight increase in metric scores, although it is not statistically significant.
For all but one other configuration, we see a statistically significant increase in metric scores for both METEOR and USE.
We see a similar pattern for the 44k output vocabulary set as the metric scores change insignificantly for $\epsilon$=0.001 and $\epsilon$=0.003 but then achieve high improvement for all other configurations.
Note that we do not recommend any particular model for the problem of source code summarization.
Rather we demonstrate how all existing models benefit in performance from adding label smoothing.

\begin{table}[!t]
	\small
	\caption{\small{Scores of astattendgru-fc for different $\epsilon$. Output vocabulary size for the top table is 10k and for the bottom is 44k.}}
	\vspace{-.1cm}
	\setlength\tabcolsep{5pt}
	\begin{tabular}{p{0.6cm}|p{0.6cm}p{0.6cm}p{0.6cm}|p{1cm}p{1cm}}
		& \multicolumn{3}{c|}{\emph{metric scores}} & \multicolumn{2}{c}{\emph{t-stat, p-value}} \\
		$\epsilon$ 	& \multicolumn{1}{c}{\small{METEOR}} 		& \multicolumn{1}{c}{\small{USE}} 	& \multicolumn{1}{c|}{\small{BLEU}}
		& \multicolumn{1}{c}{\small{METEOR}}  & \multicolumn{1}{c}{\small{USE}} \\  \cline{1-6}
		0	        & \multicolumn{1}{c}{33.09} & \multicolumn{1}{c}{50.05} & \multicolumn{1}{c|}{18.76}
		& \multicolumn{1}{c}{-} 	& \multicolumn{1}{c}{-} \\
		0.001       & \multicolumn{1}{c}{32.74} & \multicolumn{1}{c}{49.53} & \multicolumn{1}{c|}{18.66} 			
		& \multicolumn{1}{r}{-2.26, \textcolor{white}{$<$}0.02} 	& \multicolumn{1}{r}{-3.24, $<$0.01} \\
		0.003       & \multicolumn{1}{c}{33.32} & \multicolumn{1}{c}{50.23} & \multicolumn{1}{c|}{19.07} 			
		& \multicolumn{1}{r}{1.40, \textcolor{white}{$<$}0.16} 	& \multicolumn{1}{r}{1.04, \textcolor{white}{$<$}0.30} \\
		0.007       & \multicolumn{1}{c}{33.62} & \multicolumn{1}{c}{50.60} & \multicolumn{1}{c|}{19.08} 			
		& \multicolumn{1}{r}{3.26, $<$0.01} 	& \multicolumn{1}{r}{3.25, $<$0.01} \\
		0.02       	& \multicolumn{1}{c}{33.49} & \multicolumn{1}{c}{50.16} & \multicolumn{1}{c|}{19.18} 			
		& \multicolumn{1}{r}{2.40, \textcolor{white}{$<$}0.02} 	& \multicolumn{1}{r}{0.64, \textcolor{white}{$<$}0.52} \\
		0.05	    & \multicolumn{1}{c}{33.50} & \multicolumn{1}{c}{50.81} & \multicolumn{1}{c|}{19.41} 			
		& \multicolumn{1}{r}{2.47, \textcolor{white}{$<$}0.01} 	& \multicolumn{1}{r}{4.56, $<$0.01} \\
		0.10       	& \multicolumn{1}{c}{33.66} & \multicolumn{1}{c}{50.91} & \multicolumn{1}{c|}{19.38} 			
		& \multicolumn{1}{r}{3.47, $<$0.01} 	& \multicolumn{1}{r}{5.10, $<$0.01} \\
		0.25       	& \multicolumn{1}{c}{33.80} & \multicolumn{1}{c}{50.88} & \multicolumn{1}{c|}{19.54} 			
		& \multicolumn{1}{r}{4.21, $<$0.01} 	& \multicolumn{1}{r}{4.76, $<$0.01} \\
		0.40        & \multicolumn{1}{c}{33.98} & \multicolumn{1}{c}{51.22} & \multicolumn{1}{c|}{19.58} 			
		& \multicolumn{1}{r}{5.48, $<$0.01} 	& \multicolumn{1}{r}{7.06, $<$0.01}
	\end{tabular}
	\vspace{0.2cm}
	
	\begin{tabular}{p{0.6cm}|p{0.6cm}p{0.6cm}p{0.6cm}|p{1cm}p{1cm}}
		& \multicolumn{3}{c|}{\emph{metric scores}} & \multicolumn{2}{c}{\emph{t-stat, p-value}} \\
		$\epsilon$ 	& \multicolumn{1}{c}{\small{METEOR}} 		& \multicolumn{1}{c}{\small{USE}} 	& \multicolumn{1}{c|}{\small{BLEU}}
		& \multicolumn{1}{c}{\small{METEOR}}  & \multicolumn{1}{c}{\small{USE}} \\  \cline{1-6}
		0       & \multicolumn{1}{c}{32.69} 	& \multicolumn{1}{c}{49.87} 		& \multicolumn{1}{c|}{18.65} 
		& \multicolumn{1}{c}{-} 	& \multicolumn{1}{c}{-} \\
		0.001       & \multicolumn{1}{c}{33.03} 	& \multicolumn{1}{c}{49.54} 		& \multicolumn{1}{c|}{18.96} 			
		& \multicolumn{1}{r}{2.06, \textcolor{white}{$<$}0.04} 	& \multicolumn{1}{r}{-1.97, \textcolor{white}{$<$}0.05} \\
		0.003       & \multicolumn{1}{c}{32.89} 	& \multicolumn{1}{c}{50.19} 		& \multicolumn{1}{c|}{18.85} 			
		& \multicolumn{1}{r}{1.24, \textcolor{white}{$<$}0.21} 	& \multicolumn{1}{r}{1.93, \textcolor{white}{$<$}0.05} \\
		0.007       & \multicolumn{1}{c}{33.36} 	& \multicolumn{1}{c}{50.57} 		& \multicolumn{1}{c|}{19.07} 			
		& \multicolumn{1}{r}{4.13, $<$0.01} 	& \multicolumn{1}{r}{4.31, $<$0.01} \\
		0.02       	& \multicolumn{1}{c}{33.69} 	& \multicolumn{1}{c}{50.73} 		& \multicolumn{1}{c|}{19.24} 			
		& \multicolumn{1}{r}{6.15, $<$0.01} 	& \multicolumn{1}{r}{5.23, $<$0.01} \\
		0.05	    & \multicolumn{1}{c}{33.58} 	& \multicolumn{1}{c}{50.92} 		& \multicolumn{1}{c|}{19.39} 			
		& \multicolumn{1}{r}{5.42, $<$0.01} 	& \multicolumn{1}{r}{6.37, $<$0.01} \\
		0.10       	& \multicolumn{1}{c}{33.75} 	& \multicolumn{1}{c}{51.37} 		& \multicolumn{1}{c|}{19.35} 			
		& \multicolumn{1}{r}{6.34, $<$0.01} 	& \multicolumn{1}{r}{8.77, $<$0.01} \\
		0.25       	& \multicolumn{1}{c}{34.27} 	& \multicolumn{1}{c}{51.95} 		& \multicolumn{1}{c|}{19.75} 			
		& \multicolumn{1}{r}{9.76, $<$0.01} 	& \multicolumn{1}{r}{12.77, $<$0.01} \\
		0.40       & \multicolumn{1}{c}{33.84} 	& \multicolumn{1}{c}{51.21} 		& \multicolumn{1}{c|}{19.46} 			
		& \multicolumn{1}{r}{7.21, $<$0.01} 	& \multicolumn{1}{r}{8.18, $<$0.01}
	\end{tabular}
\vspace{-0.4cm}
\label{tab:rq2astattendgrufc}
\end{table}
\begin{figure*}[!b]
	\vspace{-0.4cm}
	\minipage{0.32\textwidth}
	\includegraphics[width=\linewidth, height=6cm]{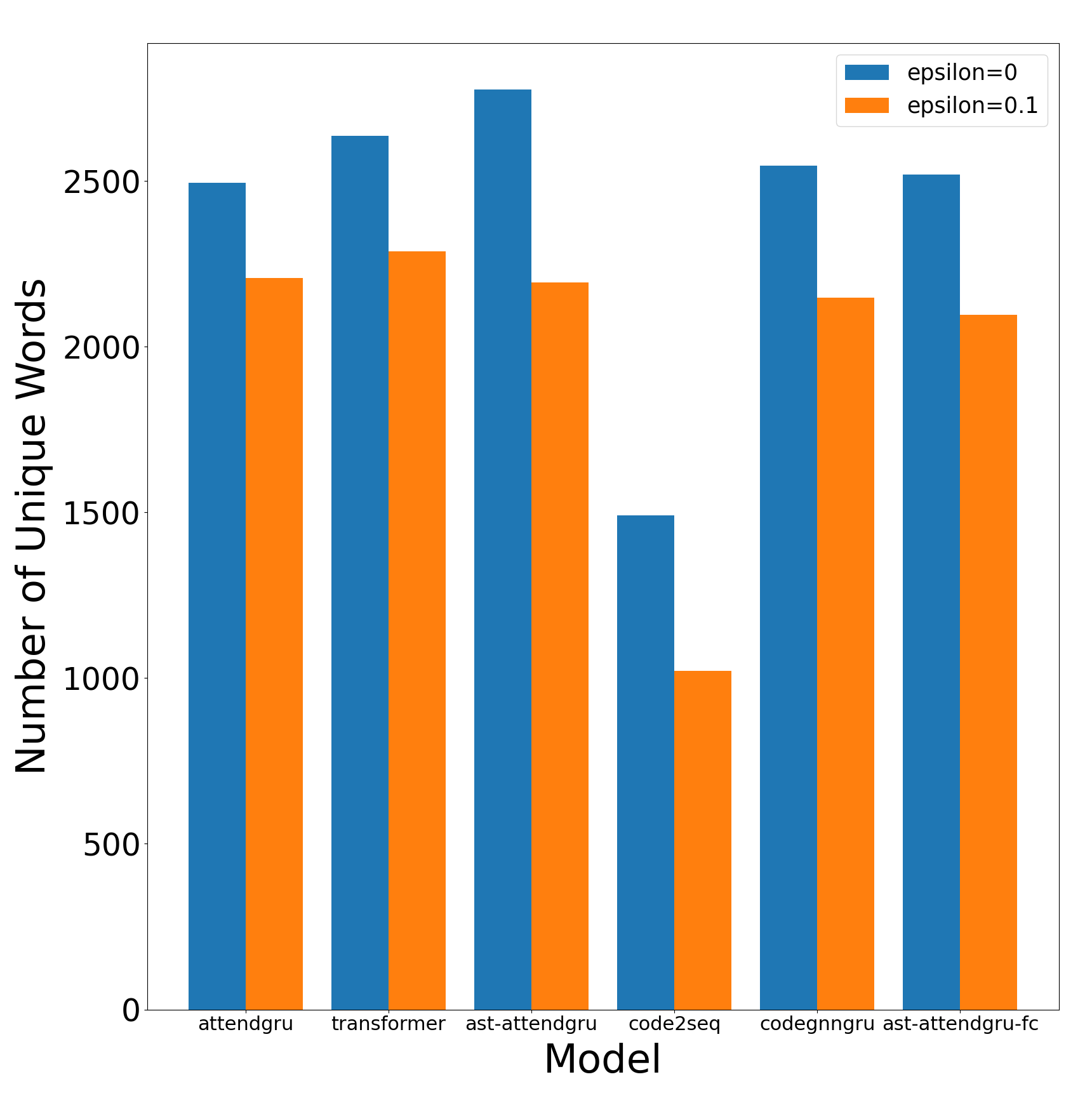}
	%\caption{A really Awesome Image}\label{fig:rq1q90}
	\endminipage\hfill
	%\vspace{-0.5cm}
	\minipage{0.32\textwidth}
	\includegraphics[width=\linewidth, height=6cm]{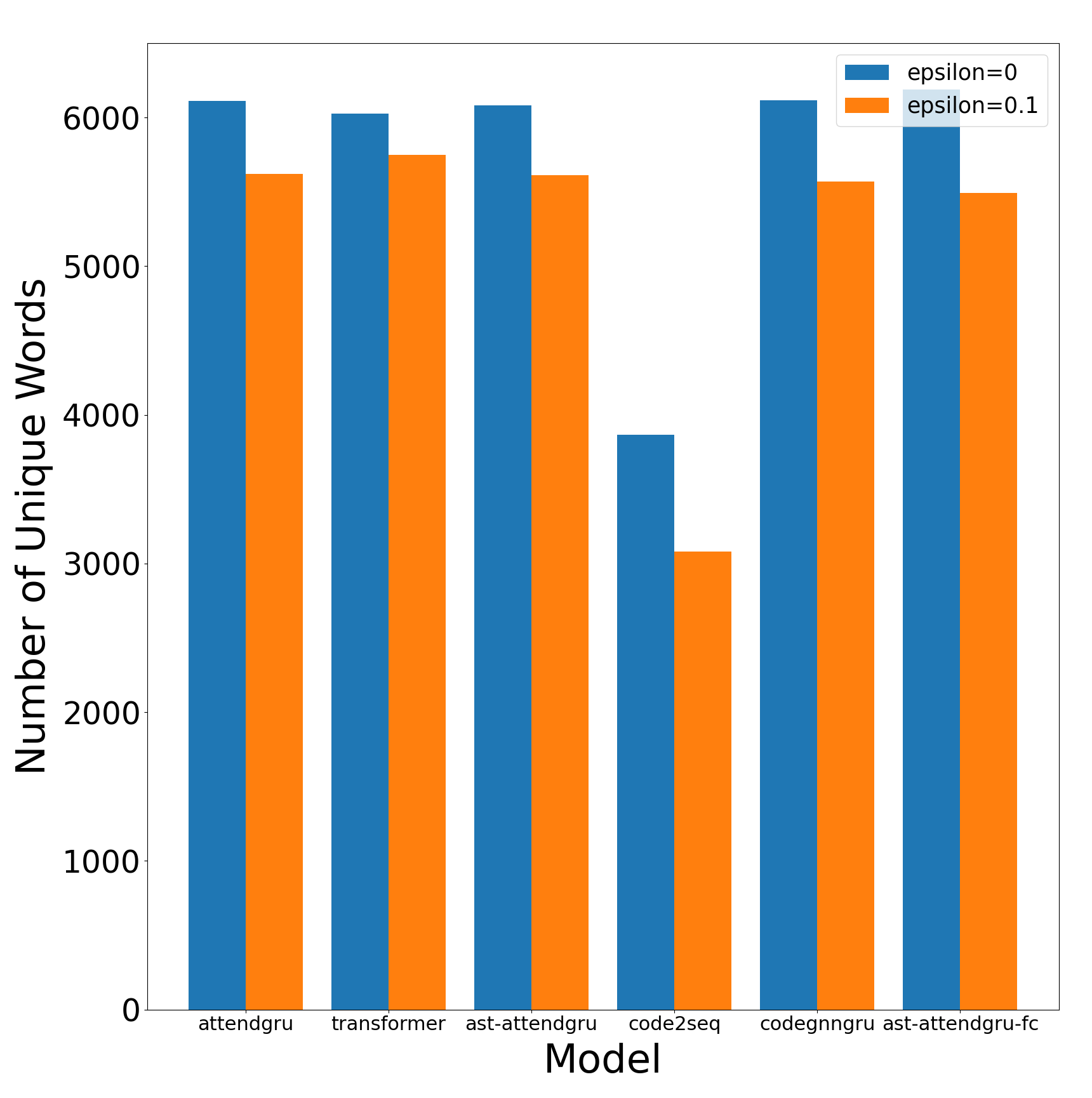}
	%\caption{A really Awesome Image}\label{fig:rq1q75}
	\endminipage\hfill
	%\vspace{-0.5cm}
	\minipage{0.32\textwidth}
	\includegraphics[width=\linewidth, height=6cm]{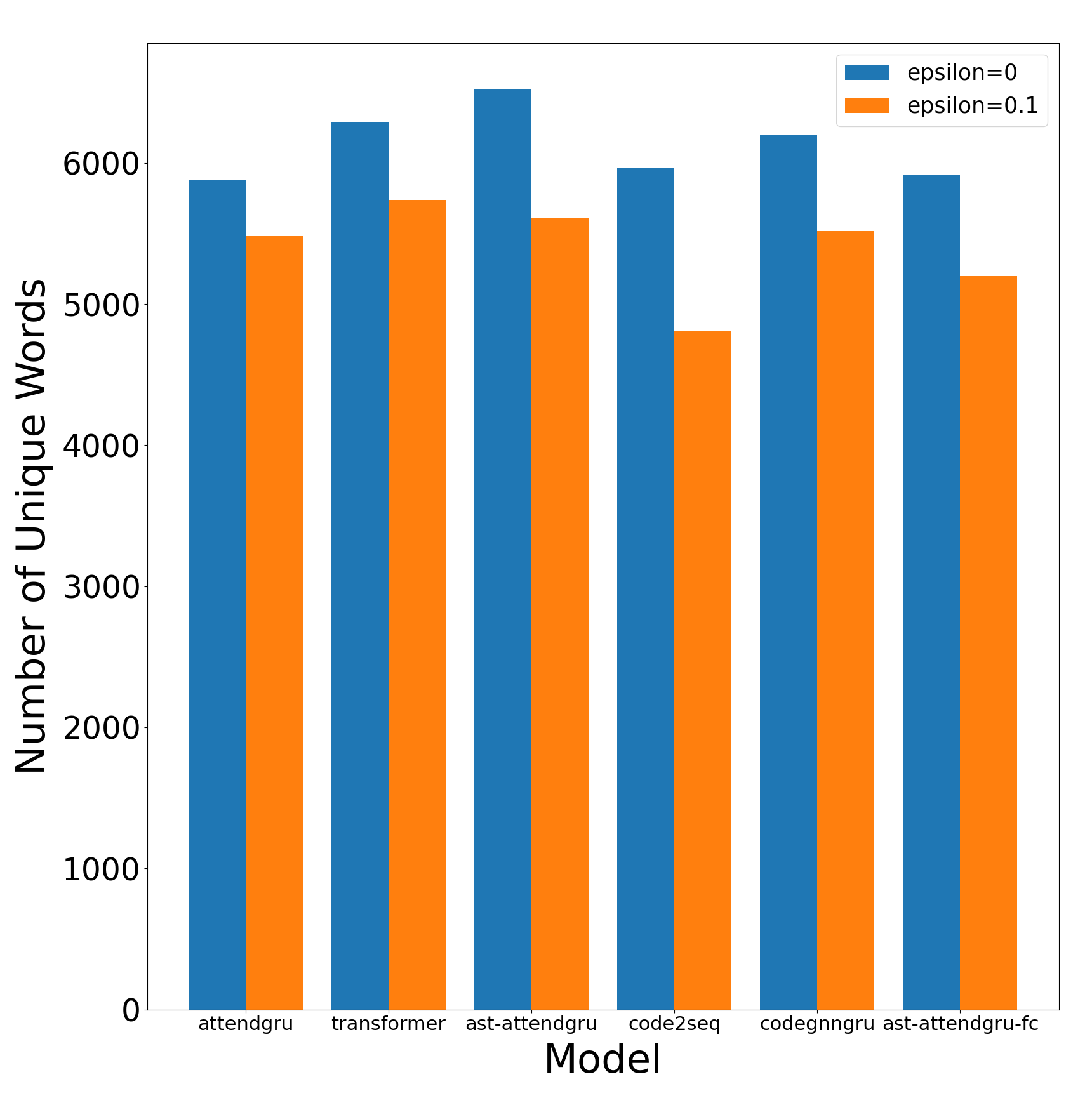}
	\endminipage
	%\captionsetup{justification=centering}
	\vspace{-0.2cm}
	\caption{Comparison of word diversity for baselines with and without label smoothing ($\epsilon$=0.10). Blue indicates baseline total number of unique words. Orange indicates total number of unique words for identical model with label smoothing. The left figure is for the Java-q90 dataset, the center figure is for the C/C++-q75 dataset, and the right figure is for the full Java dataset.}
	\label{fig:rq3rq1d}
\end{figure*}

\section{RQ3 - Vocabulary Diversity}

\subsubsection*{Methodology}

To answer RQ3, we look at the output vocabulary distribution for all the models in RQ1 and RQ2.
We find the total number of words predicted per model as well as the total number of unique words per prediction set.
This helps us identify the average word frequency in the prediction set.
We further try to identify how label smoothing affects model performance on the action word prediction problem.
We re-implement each model discussed in Section~\ref{sec:baselines} to only predict the action word.
Related work predict the top-10 or top-40 most commonly occurring action words and bucket the rest under ``other.''
We try to predict all the action words instead to see how the diversity of output vocabulary changes with label smoothing.
We train each model for the Java-q90 and C/C++-q75 dataset.
We train each baseline three times: once without label smoothing, once with $\epsilon$=0.1 and once with $\epsilon$=0.4.
We train each configuration for 10 epochs and choose the model with the highest validation accuracy score for our comparison (standard practice in related work).
We then evaluate the performance of these models using precision, recall and f1-score.
This is because action word prediction is a multi-class classification problem, while BLEU, METEOR and USE+c are more suited for full sentence prediction.

\subsubsection*{Key Findings}

Our key finding in answering RQ3 is that label smoothing decreases word diversity in output vocabulary.
As we increase $\epsilon$, the total number of words in the output prediction remains similar but the total number of unique words decreases, thus increasing the average output word frequency.
We also find that the action word prediction is unaffected by label smoothing.
Therefore, we conclude that the improvement in full model performance perceived in RQ1 must be coming from the rest of the sentence.

\begin{figure}[!b]
	\vspace{-0.7cm}
	\includegraphics[width=\linewidth, height=8cm]{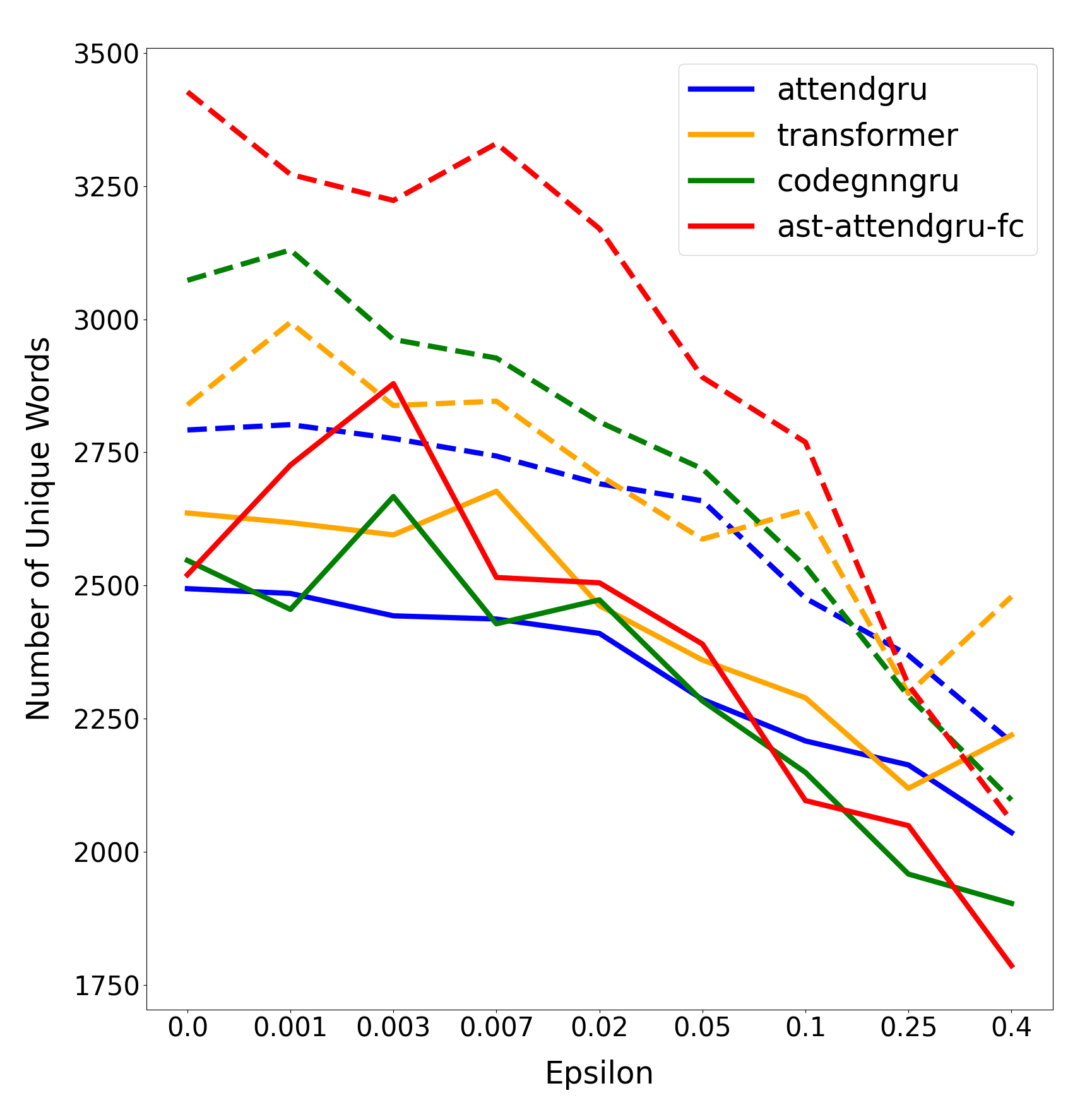}
	\vspace{-0.6cm}
	\caption{Comparison of word diversity for baselines with different values of $\epsilon$. The solid lines are trained with 10k output vocabulary and the dashed lines are trained with 44k output vocabulary. Note that the x-axis is logarithmic.}
	\label{fig:rq3rq2d}
	\vspace{-0.2cm}
\end{figure}

Figure~\ref{fig:rq3rq1d} shows how the number of unique words change with the addition of label smoothing for all the models in RQ1.
The blue bar represents models without label smoothing.
The orange bar represents the same models with label smoothing ($\epsilon$=0.1).
As the figures show, the total number of unique words decrease for all baselines, for all datasets.
%For the Java-q90 dataset (left) the number of unique words decreases substantially for each model, with the smallest decrease for attendgru (13.0\%) and the largest decrease for code2seq (46\%).
%We see a similar trend with the C/C++-q75 dataset (center) with a 4.8\% (transformer) to 25.5\% (code2seq) decrease in word diversity, and the full java dataset (right) with 7.3\% (attendgru) to 23.9\% (code2seq) decrease in word diversity.
To further study this trend, we run the same word diversity study on the models in RQ2.
Figure~\ref{fig:rq3rq2d} shows how the number of unique words change with changing values of the $\epsilon$.
A clear pattern emerges regardless of output vocabulary size (solid for 10k output vocabulary size and dashed for 44k output vocabulary size): The total number of unique words decrease steadily as we increase the value of $\epsilon$.
It is interesting to note that ast-attendgru-fc experiences highest improvement in performance AND the highest decrease in the number of unique words (29.1\% for 10k output and 39.9\% for 44k output).

Both figures suggest that label smoothing improves model performance by predicting commonly occurring words more often, thus reducing word diversity.
This explains how label smoothing achieves generalization: it avoids predicting rare words from the output distribution to reduce model loss.

To further narrow down how label smoothing improves comment generation, we test its effect on action word prediction problem.
Haque~\emph{et al.} showed that the action word prediction is a crucial stepping stone problem in source code summarization and models that predict action words poorly also predict the rest of the comment sentence poorly.
We introduce label smoothing to the action word prediction problem to see if it improves model performance while reducing the number of unique action words predicted.

\begin{table}[!t]
	\small
	\centering
	\caption{Results for RQ3 on Java q90 dataset.}
	\vspace{-0.1cm}
	\label{tab:rq3q90}
	\begin{tabular}{p{2.2cm}|p{0.2cm}p{0.2cm}p{0.2cm}|p{0.5cm}p{0.5cm}p{0.5cm}|p{0.5cm}p{0.5cm}p{0.5cm}}
					& \multicolumn{3}{c|}{\emph{$\epsilon=0$}}  & \multicolumn{3}{c|}{\emph{$\epsilon=0.1$}} & \multicolumn{3}{c}{\emph{$\epsilon=0.4$}} \\
	Model Name 		& \multicolumn{1}{c}{P} 		& \multicolumn{1}{c}{R} 		& \multicolumn{1}{c|}{F1} 
					& \multicolumn{1}{c}{P} 		& \multicolumn{1}{c}{R} 		& \multicolumn{1}{c|}{F1}
					& \multicolumn{1}{c}{P} 		& \multicolumn{1}{c}{R} 		& \multicolumn{1}{c}{F1} \\ \cline{1-10} 
	Attendgru       & \multicolumn{1}{c}{0.43} 		& \multicolumn{1}{c}{0.51} 		& \multicolumn{1}{c|}{0.45} 
					& \multicolumn{1}{c}{0.43} 		& \multicolumn{1}{c}{0.51} 		& \multicolumn{1}{c|}{0.45}			
					& \multicolumn{1}{c}{0.44} 		& \multicolumn{1}{c}{0.52} 		& \multicolumn{1}{c}{0.46}  \\
	Transformer    	& \multicolumn{1}{c}{0.45} 		& \multicolumn{1}{c}{0.50} 		& \multicolumn{1}{c|}{0.45}
					& \multicolumn{1}{c}{0.45} 		& \multicolumn{1}{c}{0.50} 		& \multicolumn{1}{c|}{0.44} 			
					& \multicolumn{1}{c}{0.46} 		& \multicolumn{1}{c}{0.52} 		& \multicolumn{1}{c}{0.46}	\\
	AstAttendgru 	& \multicolumn{1}{c}{0.41} 		& \multicolumn{1}{c}{0.49} 		& \multicolumn{1}{c|}{0.43} 
					& \multicolumn{1}{c}{0.42} 		& \multicolumn{1}{c}{0.50} 		& \multicolumn{1}{c|}{0.43}			
					& \multicolumn{1}{c}{0.43} 		& \multicolumn{1}{c}{0.50} 		& \multicolumn{1}{c}{0.44} \\
	Code2Seq        & \multicolumn{1}{c}{0.41} 		& \multicolumn{1}{c}{0.50} 		& \multicolumn{1}{c|}{0.44} 
					& \multicolumn{1}{c}{0.41} 		& \multicolumn{1}{c}{0.50} 		& \multicolumn{1}{c|}{0.44}			
					& \multicolumn{1}{c}{0.42} 		& \multicolumn{1}{c}{0.51} 		& \multicolumn{1}{c}{0.44} \\
	CodeGNNGRU      & \multicolumn{1}{c}{0.43} 		& \multicolumn{1}{c}{0.51} 		& \multicolumn{1}{c|}{0.45} 
					& \multicolumn{1}{c}{0.43} 		& \multicolumn{1}{c}{0.51} 		& \multicolumn{1}{c|}{0.44}			
					& \multicolumn{1}{c}{0.43} 		& \multicolumn{1}{c}{0.51} 		& \multicolumn{1}{c}{0.45} \\
	AstAttendgru-fc & \multicolumn{1}{c}{0.44} 		& \multicolumn{1}{c}{0.52} 		& \multicolumn{1}{c|}{0.46} 
					& \multicolumn{1}{c}{0.45} 		& \multicolumn{1}{c}{0.52} 		& \multicolumn{1}{c|}{0.45}			
					& \multicolumn{1}{c}{0.45} 		& \multicolumn{1}{c}{0.52} 		& \multicolumn{1}{c}{0.46} \\
	\end{tabular}
\end{table}

\begin{table}[!t]
	\small
	\centering
	\vspace{0.2cm}
	\caption{Results for RQ3 on C/C++ q75 dataset.}
	\vspace{-0.1cm}
	\label{tab:rq3q75}
	\begin{tabular}{p{2.2cm}|p{0.2cm}p{0.2cm}p{0.2cm}|p{0.5cm}p{0.5cm}p{0.5cm}|p{0.5cm}p{0.5cm}p{0.5cm}}
		& \multicolumn{3}{c|}{\emph{$\epsilon=0$}}  & \multicolumn{3}{c|}{\emph{$\epsilon=0.1$}} & \multicolumn{3}{c}{\emph{$\epsilon=0.4$}} \\
		Model Name 		& \multicolumn{1}{c}{P} 		& \multicolumn{1}{c}{R} 		& \multicolumn{1}{c|}{F1} 
		& \multicolumn{1}{c}{P} 		& \multicolumn{1}{c}{R} 		& \multicolumn{1}{c|}{F1}
		& \multicolumn{1}{c}{P} 		& \multicolumn{1}{c}{R} 		& \multicolumn{1}{c}{F1} \\ \cline{1-10} 
		Attendgru       & \multicolumn{1}{c}{0.69} 		& \multicolumn{1}{c}{0.71} 		& \multicolumn{1}{c|}{0.69} 
		& \multicolumn{1}{c}{0.70} 		& \multicolumn{1}{c}{0.71} 		& \multicolumn{1}{c|}{0.69}			
		& \multicolumn{1}{c}{0.71} 		& \multicolumn{1}{c}{0.72} 		& \multicolumn{1}{c}{0.70}  \\
		Transformer    	& \multicolumn{1}{c}{0.71} 		& \multicolumn{1}{c}{0.72} 		& \multicolumn{1}{c|}{0.70}
		& \multicolumn{1}{c}{0.72} 		& \multicolumn{1}{c}{0.73} 		& \multicolumn{1}{c|}{0.71} 			
		& \multicolumn{1}{c}{0.72} 		& \multicolumn{1}{c}{0.73} 		& \multicolumn{1}{c}{0.71}	\\
		AstAttendgru 	& \multicolumn{1}{c}{0.69} 		& \multicolumn{1}{c}{0.69} 		& \multicolumn{1}{c|}{0.67} 
		& \multicolumn{1}{c}{0.70} 		& \multicolumn{1}{c}{0.70} 		& \multicolumn{1}{c|}{0.68}			
		& \multicolumn{1}{c}{0.71} 		& \multicolumn{1}{c}{0.71} 		& \multicolumn{1}{c}{0.68} \\
		Code2Seq        & \multicolumn{1}{c}{0.62} 		& \multicolumn{1}{c}{0.63} 		& \multicolumn{1}{c|}{0.61} 
		& \multicolumn{1}{c}{0.62} 		& \multicolumn{1}{c}{0.62} 		& \multicolumn{1}{c|}{0.60}			
		& \multicolumn{1}{c}{0.62} 		& \multicolumn{1}{c}{0.62} 		& \multicolumn{1}{c}{0.60} \\
		CodeGNNGRU      & \multicolumn{1}{c}{0.69} 		& \multicolumn{1}{c}{0.70} 		& \multicolumn{1}{c|}{0.68} 
		& \multicolumn{1}{c}{0.70} 		& \multicolumn{1}{c}{0.71} 		& \multicolumn{1}{c|}{0.69}			
		& \multicolumn{1}{c}{0.70} 		& \multicolumn{1}{c}{0.72} 		& \multicolumn{1}{c}{0.69} \\
		AstAttendgru-fc & \multicolumn{1}{c}{0.69} 		& \multicolumn{1}{c}{0.70} 		& \multicolumn{1}{c|}{0.67} 
		& \multicolumn{1}{c}{0.70} 		& \multicolumn{1}{c}{0.71} 		& \multicolumn{1}{c|}{0.69}			
		& \multicolumn{1}{c}{0.70} 		& \multicolumn{1}{c}{0.72} 		& \multicolumn{1}{c}{0.69} \\
	\end{tabular}
\end{table}

Tables~\ref{tab:rq3q90} and~\ref{tab:rq3q75} show the effect of label smoothing on action word prediction.
We train the models under 3 conditions: $\epsilon$=0 (no label smoothing), $\epsilon$=0.1 and $\epsilon$=0.4.
We see that label smoothing has little to no effect on model performance for action word prediction.
%No model shows an improvement greater than 0.01 between $\epsilon$=0 and $\epsilon$=0.1 and greater than 0.02 between $\epsilon$=0 and $\epsilon$=0.4 for precision, recall or f1-score.
No model shows an improvement greater than 0.02 between $\epsilon$=0 and $\epsilon$=0.4 for precision, recall or f1-score.
Figure 4 shows the distribution of unique action words for each model with each label smoothing factor for the Java-q90 and C/C++-q75 dataset respectively.
We can see that there is no clear trend in the diversity of action words with changing values of $\epsilon$ for either dataset: the number of unique action words decrease in some cases and increase in others.

Two key findings can be derived from this: 1) The performance improvement we see in RQ1 and RQ2 is not due to the improved action word prediction, rather most likely coming from predicting the rest of the sentence.
2) Adding label smoothing does not affect the diversity of unique action words.
These findings further bolster the intuition that label smoothing improves model generalization by predicting more commonly occurring words in favor of rarer words.
This explains why we recommend the use of $\epsilon$=0.1 instead of 0.4 in RQ2.
While 0.4 squeezes out the best performance, most models demonstrate statistically significant improvement at 0.1 as well.
Further improvement comes at the cost of reduced word diversity.% which is often undesirable as this generates boring comments.

\begin{figure}[!t]
	\vspace{-0.5cm}
	\raggedleft
	%\left
	\minipage{0.35\textwidth}
	\caption{Action word diversity with ($\epsilon=0.1, 0.4$), without ($\epsilon=0$) label smoothing.}
	\includegraphics[width=6.5cm, height=3.5cm]{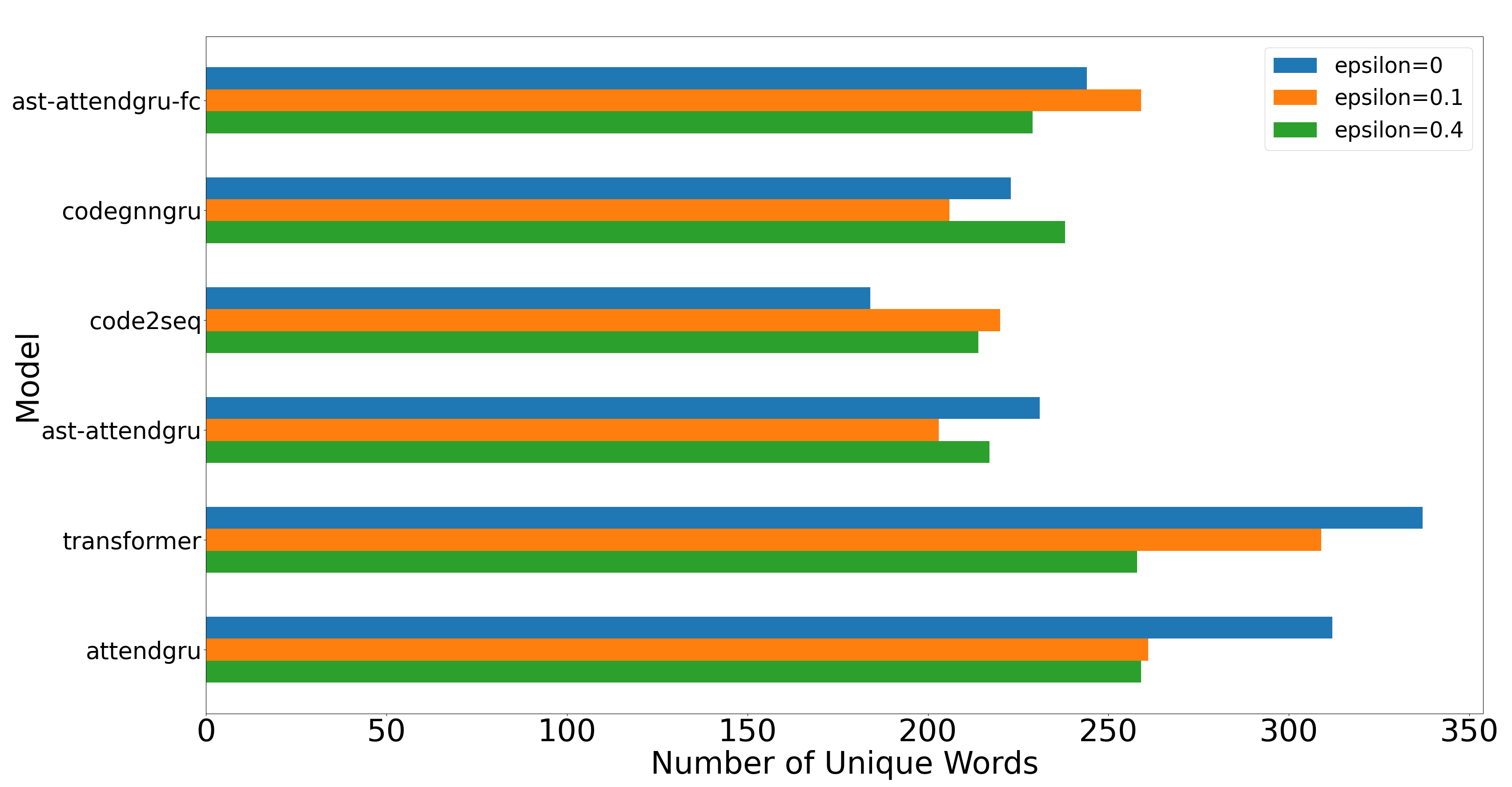}
	\endminipage\hspace{0.2cm}%\hfill
	\captionsetup{justification=centering}
	\label{fig:rq3awq90}
	\vspace{-0.4cm}
\end{figure}

\begin{figure}[!t]
	\raggedleft
	\minipage{0.35\textwidth}
	\includegraphics[width=6.5cm, height=3.5cm]{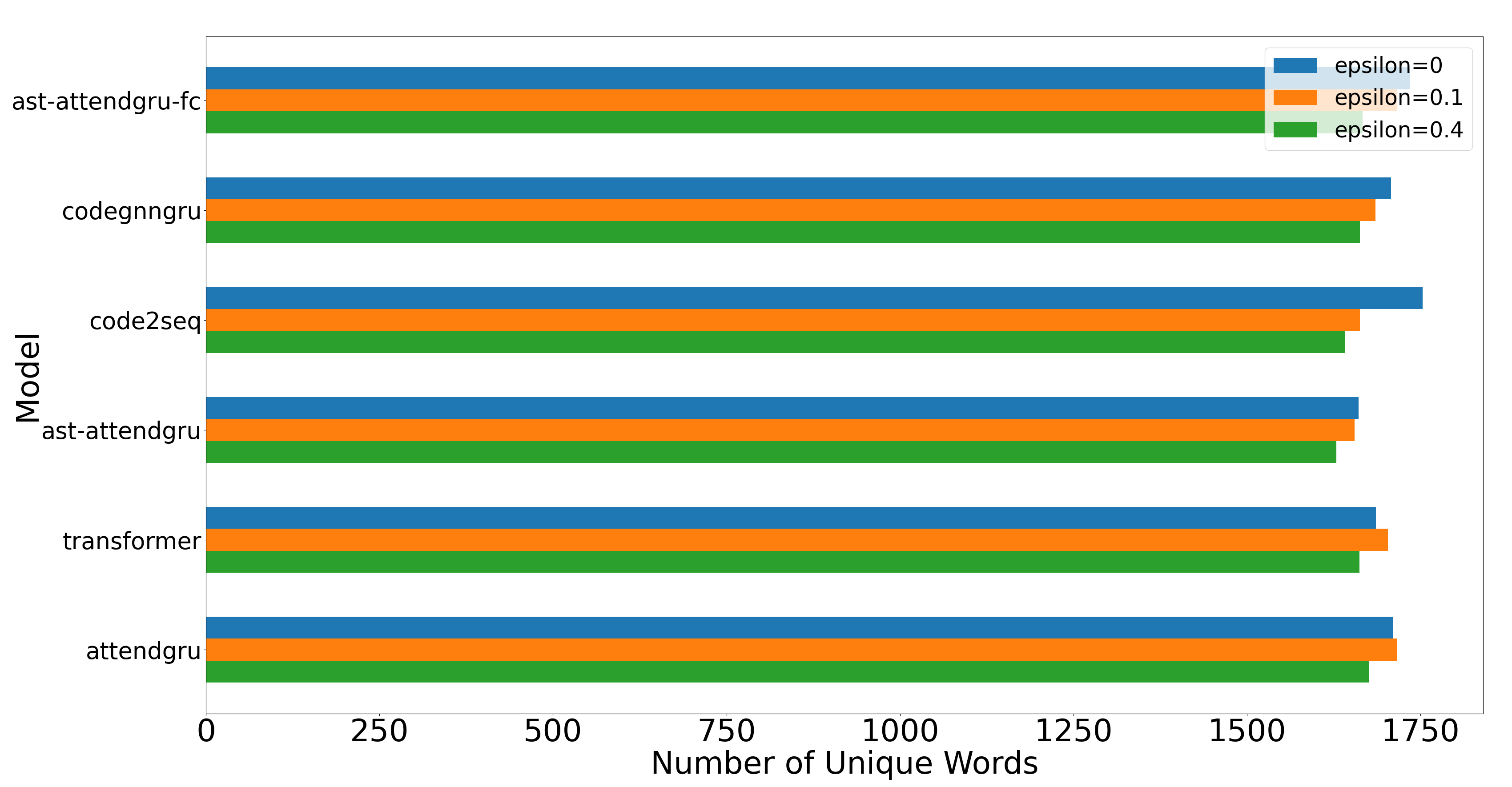}
	\captionsetup{justification=centering}
	\endminipage\hspace{0.2cm}%\hfill
	\label{fig:rq3awq75}
	\vspace{-0.4cm}
\end{figure}

%discuss figure 2 and 3
%Write numbers for figure 2 and 3

%discuss the findings of tables 9 and 10

\section{Discussion \& Conclusion}
This paper contributes to source code summarization research along three major frontiers:
\begin{itemize}
	\item[1)] We contribute a study in the use of label smoothing as a regularizer in source code summarization.
	We explore a variety of source code summarization baselines and quantitatively demonstrate that label smoothing improves model performance in terms of three different metrics.
	\item[2)] We make a recommendation on what label smoothing hyperparameter ($\epsilon$) should be used for source code summarization.
	We choose four different source code summarization models and tune the value of $\epsilon$ for best performance.
	We identify that higher values of $\epsilon$ generally lead to higher model performance.
	Our recommendation is to use $\epsilon$=0.1 for all source code summarization models.
	Our justification for this recommendation is in Section~\ref{sec:rq2_method}.
	\item[3)] We contribute new insight in the understanding of how label smoothing generalizes NLG models.
	We explore the word diversity of output vocabulary for all models trained with and without label smoothing in RQ1 and RQ2.
	We find that label smoothing reduces the number of unique words predicted in all cases.
	We further identify that label smoothing does not affect action word prediction.
%	We further add label smoothing to the action word prediction problem, and identify that label smoothing improves the prediction of the rest of the sentence, not action word.
	Thus, we posit that for subsequent word prediction in the output sentence, label smoothing improves model performance by predicting commonly occurring words more often and avoiding rare/unique words to reduce model loss.
\end{itemize}
%Prevailing evidence in other NLG work along with evidence from this paper shows that label smoothing improves model performance.
%Yet, there is still a lack of understanding as to why or how label smoothing works.
%The general intuition is that label smoothing acts as a regulizer to prevent over-fitting and make models less confident.
%However, there are opposing schools of thought as to how the internal mechanism of label smoothing helps models generalize predictions~\cite{lukasik2020does}.
%One intuition is that label smoothing makes models less certain by introducing some noise.
%The smoothed probability across the one-hot vector encourages models to predict more rare words. 
%While this exacerbates the problem of label noise, it prevents models from being over confident, thus having an overall generalization effect.
%A competing intuition is that label smoothing improves model calibration by reducing over-fitting.
%This implies that the model does not interpret noise as pattern and hence produces more generalized predictions.
%Our findings reinforce the findings by Lukasik~\emph{et al.}~\cite{lukasik2020does} who demonstrate that label smoothing reduces label noise.

%To encourage reproducibility and aid other research groups, we have released all data, source code, scripts, and tutorial information in an online appendix:

%We provide the following online appendix for reproducibility

%\section{Reproducibility}
%To encourage reproducibility and aid other research groups, we release all data and source code in the following online appendix:
Appendix:
\texttt{https://github.com/ls-scs/ls\_scs}

\section*{Acknowledgment}

This work is supported in part by NSF CCF-1452959 and CCF-1717607. Any opinions, findings, and conclusions expressed herein are the authors and do not necessarily reflect those of the sponsors.

\bibliographystyle{IEEEtran}
\bibliography{main}

\end{document}